\documentclass[a4paper, 11pt]{article}
\usepackage[paper=a4paper,margin=1in]{geometry}
\usepackage{hyperref}
\usepackage{booktabs}
\usepackage{amsmath}
\usepackage[flushleft]{threeparttable}
\usepackage{adjustbox}
\usepackage{graphicx}
\usepackage{subcaption}
\usepackage{color}

\usepackage{amsthm,amsmath,amssymb} %for theorems

\newtheorem{mydef}{Definition}

\usepackage{natbib}

\usepackage{lmodern}

\hypersetup{pdfstartview={FitV}, colorlinks, urlcolor=blue, linkcolor=blue, citecolor=blue, linktoc=page}

\usepackage{authblk}

\usepackage{setspace}

\usepackage{framed}
\begin{document}

	%% title page
	\title{\textbf{Agent-Based Model Calibration using Machine Learning Surrogates}}
	\date{\today}
	\author{\Large{Francesco Lamperti}%\textsuperscript{\textit{a},}
	\thanks{Corresponding author. Institute of Economics, Scuola Superiore Sant'Anna, Piazza Martiri della Libert\`a 33, 56127 Pisa (Italy). Email: \href{mailto:f.lamperti@santannapisa.it}{f.lamperti@santannapisa.it}.},  \Large{Andrea Roventini}\thanks{Institute of Economics, Scuola Superiore Sant'Anna (Pisa) and OFCE-Sciences Po (Nice). Email: \href{mailto:a.roventini@santannapisa.it}{a.roventini@santannapisa.it}.} \hspace{0.1cm}  and  \Large{Amir Sani}\thanks{Universit\'{e} Paris 1 Path\'{e}on-Sorbonne and CNRS. Email: \href{mailto:reachme@amirsani.com}{reachme@amirsani.com}.}
	}

	\maketitle

\vspace{1.5cm}

	\begin{abstract}
		\noindent Taking agent-based models (ABM) closer to the data is an open challenge. This paper explicitly tackles parameter space exploration and calibration of ABMs combining supervised machine-learning and intelligent sampling to build a surrogate meta-model. The proposed approach provides a fast and accurate approximation of model behaviour, dramatically reducing computation time. In that, our machine-learning surrogate facilitates large scale explorations of the parameter-space, while providing a powerful filter to gain insights into the complex functioning of agent-based models. The algorithm introduced in this paper merges model simulation and output analysis into a surrogate meta-model, which facilitates fast and efficient ABM calibration. We successfully apply our approach to the \cite{BH98} asset pricing model and to the ``Island'' endogenous growth model \citep{FD03}. Performance is evaluated against a relatively large out-of-sample set of parameter combinations, while employing different user-defined statistical tests for output analysis. The results demonstrate the capacity of machine learning surrogates to facilitate fast and precise exploration of agent-based models' behaviour over their often rugged parameter spaces.

	\end{abstract}
	
\vspace{1.5cm}	

\noindent\textbf{Keywords}: agent based model; calibration; machine learning; surrogate; meta-model.\\
\textbf{JEL codes:} C15, C52, C63. 	
	
	\newpage

	\onehalfspacing
	
	\section{Introduction}
	
	This paper proposes a novel approach to model calibration and parameter space exploration in agent-based models (ABM), combining supervised machine learning and intelligent sampling in the design of a novel surrogate meta-model. 
	
	Agent-based models deal with the study of socio-ecological systems that can be properly conceptualized through a set of micro and macro relationships. 
	One problem with this framework is that the relevant statistical properties for variables of interest are \textit{a priori} unknown, even to the modeler. Such properties emerge indeed from the repeated interactions among ecologies of heterogeneous, boundedly-rational and adaptive agents.\footnote{In the last two decades a variety of ABM have been applied to study many different issues across a broad spectrum of disciplines beyond economics and including ecology \citep{grimm2013}, health care \citep{eff12}, sociology \citep{macy2002}, geography \citep{brown2005}, bio-terrorism \citep{carley2006},  medical research \citep{An2009}, military tactics \citep{ila97} and many others. See also \citet{squaz10} for a discussion on the impact of ABM in social sciences, and \citet{Fagiolo_Roventini_2012,Fagiolo_Roventini_2016} for an assessment of macroeconomic policies in agent-based models.} As a result, the dynamic properties of the system cannot be studied analytically, the identification of causal mechanisms is not always possible and interactions give rise to the emergence of relationships that cannot simply be deduced by aggregating those of micro variables (\citealp{And72}, \citealp{TJ06},  \citealp{Graz12}, \citealp{GK12}). This raises the issue of finding appropriate tools to investigate the emergent behavior of the model with respect to different parameter settings, random seeds, and initial conditions \citep[see also][]{lee15}. Once this search is successful, one can safely move to calibration, validation and, finally, employ the model for policy exercises \citep[more on that in][]{Fagiolo_Roventini_2016}. Unfortunately, this procedure is hardly implementable in practice,  notably due to large computation times.
	
	Indeed, many ABMs simulate the evolution of a complex system using many parameters and a relatively large number of time steps. In a calibration setting, this rich expressiveness results in a ``curse of dimensionality'' that lends to an exponential number of critical points along the parameter space, with multiple local maxima, minima and saddle points, which negatively impact the performance of gradient-based search procedures. Exploring model behaviour through all possible parameter combinations (a full factorial exploration) is practically impossible even for small models. Budgetary constraints also restrict our use from multi-objective optimization procedures, such as multimodel optimization or niching \citep[for a review, see e.g.][]{li2013benchmark,wong2015evolutionary}, and kriging-based procedures due to the large number of evaluations required for these procedures to converge to meaningful interpretations of the model parameter space. \footnote{For example, consider a model with $5$ parameters and assume that a single evaluation of the ABM requires $5$ seconds on a single compute core (CPU). If one discretizes the parameter space by splitting each dimension into $10$ intervals, $10^5$ evaluations would require approximately $6$ CPU days to explore. With a finer partition of of say $15$ intervals, $10^{15}$ evaluations would roughly require $1.5$ months, and $20$ intervals would require $6$ months. Adding a sixth parameter would require more than 10 years.} However, if a model is to be useful for policy makers, it must provide timely and accurate insights into the problem. As a result, for computationally expensive models such as ABMs to provide practical insights with their rich expressiveness, they must be efficiently calibrated on a limited budget of evaluations.	
	
	Traditionally, three computationally expensive steps are involved in ABM calibration; running the model, measuring calibration quality and locating parameters of interest. As remarked in \cite{Rich15}, such steps account for more than half of the time required to estimate ABMs, even for extremely simple models. Recently, kriging (also known as Gaussian processes) has been employed to build surrogate meta-models of ABMs \citep{Salle14,Dosietal16,Dosi16,Dosietal17,Bar16} to facilitate parameter space exploration and sensitivity analyses. However, kriging cannot be reasonably applied to large scale models with more than 20 parameters even in the linear time extensions proposed in \citet{wilson2015thoughts} and \citet{herlands2015scalable}. Moreover, the smooth surfaces produced by kriging meta-models do not provide an accurate approximation of the rugged parameter spaces characteristic of most ABMs. 
	
In this paper, we explicitly tackle the problem of efficiently exploring the complex parameter space of agent-based models by employing an efficient, adaptive, gradient-free search over the parameter space. The proposed approach exploits both labeled and unlabeled parameter combinations in a semi-supervised manner to build a fast, efficient, \textit{machine-learning surrogate} mapping a statistic, based on a user-defined measure of fit, and a specific parameterization of the ABM. This procedure results in a dramatic reduction in computation time, while providing an accurate surrogate of the original ABM. This surrogate can then be employed for detailed exploration of the possibly \textit{wild} parameter space. Moreover, we move towards calibration by identifying parameter combinations that allow the ABM to match user-desired properties.\footnote{The interested reader might want to look at \cite{Sander} for a broad discussion on possible applications of machine learning algorithms to agent based modelling.}
	
	Surrogate meta-models are traditionally employed to approximate or emulate computationally costly experiments or simulation models of complex physical phenomena \cite[see][]{Book99}. In particular, surrogates provide a proxy that can be exploited for fast parameter-space exploration and model calibration. Given their speed advantage, surrogates are regularly exploited to locate promising calibration values and gain rapid intuition over a model. Note that the objective is not to return a single optimal parameter, but all parametrizations that positively identify the ABM with user-desired behaviour. Accordingly, if the surrogate approximation error is small, it can be interpreted as an efficient and reasonably good replacement for the original ABM during parameter space exploration and calibration.
	
	Our approach to learning a surrogate occurs over multiple rounds. First, a large ``pool'' of unlabelled parametrizations are drawn using a standard sampling routine, such as quasi-random Sobol sampling. Next, a very small subset of the pool is randomly drawn without replacement for evaluation in the ABM, making sure to have at least one example of the user-desired behaviour. These points are ``labelled'' according to the statistic measured on the output generated by the ABM and act as a ``seed'' set of samples to initialize the surrogate model learned in the first round. This first surrogate is then exploited to predict the label for unlabelled points remaining in the pool. Another very small subset of points are drawn from the pool for evaluation in the agent-based model. Then, over multiple rounds, this process is repeated until a specified budget of evaluations is achieved. In each round, the surrogate directs which unlabelled points are drawn from the pool to maximize the performance of the surrogate learned in the next round. This semi-supervised ``active'' learning procedure incrementally improves the surrogate model, while maximizing the information gained over the ABM parameter space.\footnote{In the Machine Learning jargon supervised learning refers to the task of inferring a function from labeled training data, that is, data that are assigned either a numerical value or a symbol. Semi-supervised learning indicates a setting when there is a small amount of labelled data relatively to unlabelled ones. The term active refers instead refers an algorithm that actively selects which data point to evaluate and, therefore, to label.}

    The performance of such a procedure crucially depends on the particular surrogate model used in each of the rounds. Here, we automatically tune extremely boosted gradient trees \citep[XGBoost, see][]{chen2016xgboost} as our machine-learning surrogate, through automated hyperparameter optimization \citep[see]{claesen2014easy}, to robustly manage non-linear parameter surfaces and so-called ``knife-edge'' properties characteristic of ABMs. One particular advantage of this surrogate learning algorithm over kriging is that it does not require the selection of a kernel or to set a prior in advance of the previously mentioned sampling procedure. It also avoids the problem of choosing a summary statistic and acceptance thresholds that comes with likelihood-free approximate Bayesian methods \citep{Rich15}.
	
	As illustrative examples, we apply our procedure to two well known ABMs: the asset pricing model proposed in \cite{BH98} and the endogenous growth model developed in \citet{FD03}. Despite their relative simplicity, the two models might exhibit multiple equilibria, allow different behavioural attitudes and account for a wide range of dynamics, which crucially depends on their parameters. 
	We find that our machine-learning surrogate is able to efficiently filter out combinations of parameters conveying the output of interest, assess the relative importance of models' parameters and provide an accurate approximation of the underlying ABM in a negligible amount of time. The advantages in terms of computation cost, hands-free parameter selection and ability to deal with non-linear characteristics of the ABM parameter space of our approach paves the way towards an efficient and user-friendly procedure to parameter space exploration and calibration of agent-based models. 
	
	The remaining portions of this paper are organized as follows. Section \ref{Section:literature} reviews literature on ABM calibration validation, making the case for surrogate modelling. Section \ref{sec:surrogate} presents our surrogate modelling methodology. Sections \ref{sec:BH} and \ref{sec:IS} report the results of its application to the  asset pricing model proposed in \citet{BH98} and the growth model developed in \citet{FD03} respectively. Finally, Section \ref{sec:conclusions}  concludes.
	
	%%%%%%%%%%%%%%%%%%%%%%%%%%%%%%%%%%%%%%%%%%%%%%%%
	
	\section{Calibration and validation of agent-based models: the case for surrogate modelling}\label{Section:literature}
	
	As stated in \citet{Fagiolo07} and \citet{Fagiolo_Roventini_2012,Fagiolo_Roventini_2016}, the extreme flexibility of ABMs concerning e.g. various forms of individual behaviour, interaction patterns and institutional arrangements has allowed researchers to explore the positive and normative consequences of departing from the often over-simplifying assumptions characterizing most mainstream analytical models.
	Recent years have witnessed a trend in macro and financial modeling towards more detailed and richer models, targeting a higher number of stylized facts, and claiming a strong empirical content.\footnote{See e.g. \cite{Dosi10, Dosi13, Dosi15, Caiani2016, Assenza2015} and \cite{Dawid14} on business cycle dynamics, \cite{DSK} on growth, green transitions and climate change, \cite{dawid2014economic} on regional convergence and \cite{HFT15} on financial markets.  The surveys in \citet{Fagiolo_Roventini_2012,Fagiolo_Roventini_2016} provides a more exhaustive list.}
	
	A common theme informing both theoretical analysis and methodological
	research concerns the relationships between agent-based models and real-world data. Recently, many studies have addressed the problem of estimating and calibrating ABMs. As stated by \cite{Che12}, ABMs need to move from stage I, i.e. the capability to grow stylized facts in a qualitative sense, to stage II, where appropriate parameter values are selected according to sound econometric techniques. In those cases where the model is sufficiently simple and well behaved, one can derive a closed form solution and estimate the distribution parameters for a specific output of the model \citep[see e.g.][]{Alf05, Alf06, Bos07}. However, when complexity prevents a closed form solution, more sophisticated techniques are required. \cite{Ami08} estimates a model of financial markets with 15 parameters (but only 2 or 3 agents) by the method of moments. They report that the model has a high sensitivity to the assumptions made on the noise term and stochastic components. \cite{Gil03} and \cite{Win07} introduce an algorithm and a set of statistics leading to the construction of an objective function, which is used to estimate exchange-rate models by indirect inference, pushing them closer to the properties of real data. \citet{Fra09} refines on this framework and uses the method of simulated moments to estimate 6 parameters of an asset pricing model, while \citet{franke2012} propose a model contest for structural stochastic volatility models characterized by few parameters.\footnote{See also \cite{Graz15} and \cite{Fab12} for other applications of the same approach}  Finally, \cite{Rec15} use a simple gradient-based calibration procedure and then test the performance of the model they obtained through out of sample forecasting.
	
	A parallel stream of research has recently focusing on the development of tools to investigate the extent ABM outputs are able to approximate reality \citep[see][]{Marks13, Lamperti15, Lamperti16, Bar15, Barde16, Gue16}. Some of these contributions also offer new measures that can be used to build objective functions in the place of longitudinal moments within an estimation setting \citep[e.g. the GSL-div introduced in][]{Lamperti15}. However, a common limitation of both these calibration/estimation and validation exercises lies in their computational time, which is usually extremely high. As discussed in detail by \cite{Rich15}, simulating the model is the most computationally expensive step for all these procedures. For instance, in order to train his algorithm, \citet{Bar15} needs Monte Carlo (MC) runs each having length of about $2^{19}$ periods, and many macroeconomic ABMs might take weeks just to perform a single MC exercise of this kind. This explains why the vast majority of previous contributions employ extremely simple ABMs (few parameters, few agents, no stochastic draws) to illustrate their approach, and large macro ABMs are usually poorly validated and calibrated.
Hence, using standard statistical techniques, the number of parameters must be minimized to achieve feasible estimation. 

From a theoretical perspective, the curse of dimensionality implies that  the convergence of any estimator to the true value of a smooth function defined on a high dimensional parameter space is very slow \citep{Week95, Dem05}. Several methods have been introduced in the design of experiments literature to circumvent this problem, but the assumptions of smoothness, linearity and normality do not generally hold for ABMs \cite[see the extensive discussion in][]{lee15}.
	
	Unfortunately, recent developments in agent-based macro-economics  have led  to the development of more and more complex models, which require large sets of parameters to adequately capture the complexity of micro-founded, multi-sector and possibly multi-country phenomena \citep[see][for a recent survey]{Fagiolo_Roventini_2016}.

	In such a setting, neither direct estimations nor global sensitivity analysis \citep[often advocated as a natural approach to ABM exploration, cf.][]{moss2008alternative,thiele2014facilitating,ten2016sensitivity} seem computationally feasible. 
	
	New alternative methods must deal with two issues: reduction in computation time and the design of appropriate criteria for calibration and validation procedures. Our approach shows that such issues can be related in a meaningful way by developing a computational procedure that efficiently trains a surrogate model in order to optimize specific calibration criteria or reproducing statistical relationships between model-generated variables. Our procedure has some similarities to the one of \cite{dawid2014economic}, where penalized splines methods are employed to shortcut parameter exploration and unravel the dynamic effects of policies on the economic variables of interest. However, our method especially focuses on computational efficiency and therefore builds on two pillars: surrogate modelling and intelligent sampling.
	
	With respect to surrogate modelling, we extend recent contributions in the economic literature that use kriging to build a surrogate meta-model for ABMs \citep{Salle14, Dosi16, Bar16}. One of the primary challenges with kriging-based meta-models is that they cannot efficiently model more than a dozen parameters. This constraint forces modellers to arbitrarily fix a subset of parameters whenever the parameter space is large. Moreover, kriging relies on Gaussian processes \citep{rasmussen2006gaussian,conti2010bayesian}, which face serious difficulties when the underlying smoothness assumptions are violated. Modelling the rugged parameter space of ABMs is particularly challenging. In order to overcome these constraints, our meta-modelling approach leverages non-parametric boosted trees from the machine learning literature that do not depend on smoothness assumptions \cite[see][]{freund1996experiments, breiman1984classification}.

    Even the most advanced surrogate modelling algorithm only performs as well as the quality of labelled samples. With respect to ABMs, a labelled sample is a parameter combination and the output of the ABM given this parametrization. Batch sampling, the process of sampling a budget of samples all at once, such as in random sampling, quasi-random sampling (e.g. Sobol sampling), extensions that extend the Sobol sequence to reduce error rates (see \citealp{saltelli2010variance}) and more sophisticated procedures such as Latin-Hypercube sampling are all limited by their one-off nature to sampling. Further, ABM parameters of interest are often rare and represent a small percent of possible parametrizations. Given this \textit{imbalanced} nature of the sample and the non-negligible computation cost of evaluating ABM parameters, it makes sense to carefully select which parametrizations to evaluate, while exploiting the cheap (almost free) cost of generating unevaluated parametrizations. The problem of sequentially selecting the most informative subset of samples over multiple sampling rounds underlies \textit{active learning} \citep[see][for a survey]{settles2010active}. In particular, given a large pool of unlabelled parametrizations and a fixed evaluation budget, active learning chooses parametrizations from the pool that maximize the generalization or \textit{learning} performance of the surrogate meta-model.
	
	\section{Surrogate modelling methodology}\label{sec:surrogate}
	
	One can represent an agent-based model as a mapping $m: I \rightarrow O$ from a set of input parameters $I$ into an output set $O$. The set of parameters can be conceived as a multidimensional space spanned by the support of each parameter. The number of parameters in large macroeconomic ABMs generally range up to several dozen. The output set is generally larger, as it corresponds to time-series realizations of a very large number of micro and macro level variables. This rich set of outputs allows a qualitative validation of agent-based models based on their ability to reproduce the statistical properties of empirical data (e.g. non-stationarity of GDP, cross-correlations and relative volatilities of macroeconomic time series), as well as microeconomic distributional characteristics (e.g. distribution of firms' size, of households' income, of assets' returns). Beyond stylized facts, the quantitative validation of an agent-based model also requires the calibration/estimation of the model on a (generally small) set of aggregate variables (e.g. GDP growth rates, inflation and unemployment levels, asset returns  etc.). 
	
	Without loss of generality, we can represent this quantitative calibration as the determination of input values such that the output satisfies certain calibration conditions, coming from, e.g, a statistical test or the evaluation of a likelihood or loss functions. This is in line, for example, with the method of simulated moments \citep{Gil03, franke2012}. We consider two settings:
	
	\begin{itemize}
	\item \textbf{Binary outcome}. In this setting the calibration criterion can be considered as a function, $v: O \to \{0,1\}$, that maps the ABM output to a binary variable that takes $1$ if a certain property of the output (or set of properties) is found, and $0$ otherwise. For example, a property that one might want a financial ABM to match is the presence of excess kurtosis in the distribution of returns. This setting leads to what is referred in the machine learning literature as a \textit{classification} problem.
	
	\item \textbf{Real-valued outcome}. In this setting the calibration criterion can be considered as a function, $v: O \to \mathcal{R}$, that maps the ABM output to a real valued number providing a quantitative assessment of a certain property of the model. For example, one might want to compute excess kurtosis of simulated data and then compare it to the one obtained from real data. This setting leads to what is referred in the machine learning literature as a \textit{regression} problem.
	\end{itemize}
	
	To keep consistency with the machine learning terminology, we say that function $v$ assigns a \textit{label} to the parameter vector $x$. Obviously, one would like to find the set of input parameters $x \in I$ such that their labels indicate that a chosen condition is met. More formally, we say that $C$ is the set of labels indicating that the condition is satisfied. For example, in the case of binary outcome we can say that $C=\{1\}$, which indicates that the chosen property is observed; in the case of real-valued outcome, assuming that $v$ expresses the distance between some statistic of the simulated and real data, one might consider $C=\{x\, :\, v(x) \leq \alpha \}$ or $C=\{\min_{x\in I_j} v(x),\,\,\,j=1,2,3,..,J\}$. The latter case reflects exactly the common calibration problem of minimizing some loss function over the parameter space with random restart to avoid ending up in local minima.
	
	\begin{mydef}
	We say that a \textbf{positive calibration} is a parameter vector $x\in I$ whose label in contained in the set $C$, i.e. $x\,:\,v(x)\in C$. By contrast, a \textbf{negative calibration}  is a parameter vector whose label is not contained in $C$.
	\end{mydef}
	
The problem now is to find all positive calibrations.	However, an intensive exploration of the input set $I$ is computationally infeasible. As emphasized above, it is crucial to drastically reduce the computation time required to identify positive calibrations.

	This paper proposes to train a surrogate model that efficiently approximates the value of $f(x)=v\circ m (x)$ using a limited number of input parameters (budget) to evaluate the true ABM. Once the surrogate is trained, it provides an efficient mean of exploring the behaviour of the ABM over the entire parameter space.\footnote{Notwithstanding its precision, the surrogate remains an approximation of the original model. We suggest the user, in any case, to identify positive calibrations and further study model's behaviour therein and in their close neighbourhoods employing the original ABM.}
	
	The surrogate training procedure requires three decisions: 
	
	\begin{enumerate}
	\item  Choosing a machine learning algorithm to act as a surrogate for the original ABM, taking care that the assumptions made by the machine learning model do not force unrealistic assumptions on the parameter space;
	\item Selecting a sampling procedure to draw samples from the parameters space in order to train the surrogate;
	\item Selecting a score or criterion that can be used to evaluate the performance of the surrogate. 
	\end{enumerate}
	
We prefer to avoid smoothness assumptions and the challenges of selecting a good prior and kernel when using a kriging-based approach \citep[see ][]{rasmussen2006gaussian, ryabko2016things}, so we propose to use extreme gradient boosted trees (XGBoost) \citep[see]{chen2016xgboost}that form a random ensemble \citep[see][]{breiman2001random} of ``boosted'' \citep[see][]{freund1990boosting,freund1996experiments} classification and regression trees (CART) \citep[see]{breiman1984classification}. This choice endows our surrogate with the ability to learn non-linear ``knife-edge'' properties, which typically characterize ABM parameter spaces. Sampling should carefully select which parametrizations of an ABM should be evaluated according to the performance of the surrogate. Here, we leverage pool-based active learning according to a pre-specified budget of evaluations\footnote{For a review of active learning, see e.g. \cite{settles2010active}.} The structure of the surrogate, active learning approach and performance criterion are detailed below.
	
	\begin{figure}[tbp]
		\centering
		\includegraphics[scale=0.425]{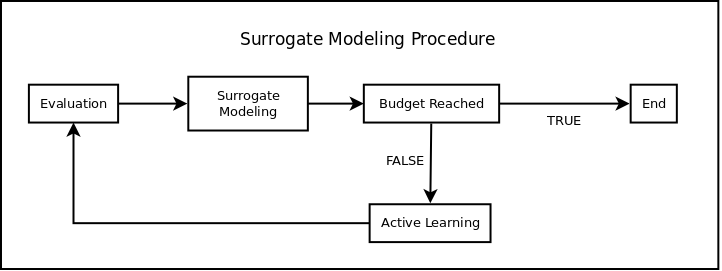}
		\caption{Surrogate modelling algorithm.}
		\label{fig:v}
	\end{figure}

	\subsection{Structure of the surrogate}

Here, we employ an iterative training procedure (see Figure \ref{fig:v}) to construct a different surrogate at each of several rounds until we approach a predefined budget of evaluations on the true ABM. At each round an additional parameter vectors is used in the iterative procedure. The budget is set in advance by the user according to a pre-determined, acceptable, computation cost of learning the surrogate. In each round, a surrogate is trained using all available parameter vectors, and their respective labels, which have been aggregated up to that round. Once the budget of evaluations is reached, the final surrogate is ready to be used for parameter space exploration.
	
\begin{figure}[ht!]
		\centering
		\includegraphics[scale=0.6]{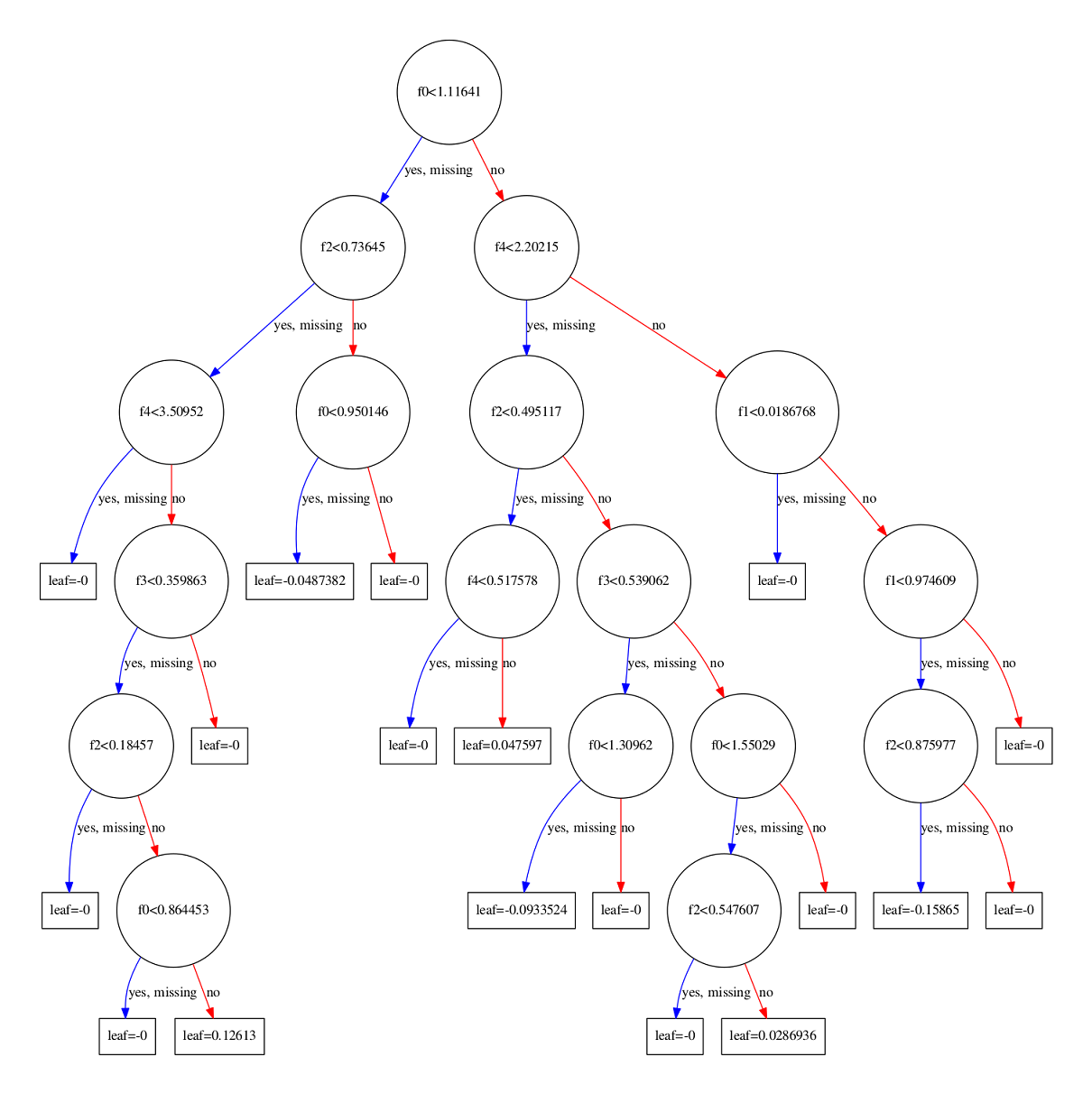}
		\caption{An example classification and regression tree (CART) used for regression. Features are labelled $f0,\dots,f4$ and nodes specify cutoff thresholds that designate the path a new parameter vector takes from the top (root) node to the final (leaf) node, which denotes the predicted calibration value. In the process of ``boosting'' CART trees to produce an ensemble, each subsequent tree increasingly focuses on the higher weighted samples. This generally results in smaller ``specialized'' trees that stick on samples that were most difficult to classify.}
		\label{fig:sample_CART}  
	\end{figure}

	Here, we rely on XGBoost \citep{chen2016xgboost} as our surrogate learning algorithm. This algorithm sequentially learns an ensemble of classification and regression trees \citep[CART, see][]{breiman1984classification}. Figure \ref{fig:sample_CART} provides an example of CART tree. Given that the CART trees are represented as functions, the gradient resulting from the ensemble of CART trees can be minimized. Weights are assigned to each of the parameter vectors and ``boosted'' in the direction of the gradient that minimizes the total loss. Boosting magnifies the importance of difficult-to-learn samples. In each of the subsequent rounds, a new tree is learned over the boosted parameter vectors, incurring an increased penalty according to the boosted weights. Accordingly, trees are learned according to the weight from the previous round. The XGBoost algorithm builds CART trees that are increasingly specialized to handle the particular subset of samples that were difficult to learn up until the current round. A common way to characterize this learning procedure is to consider it as an ensemble of ``weak'' approximations, that together construct a strong approximation \citep[see][for more details see]{freund1990boosting,freund1996experiments,chen2016xgboost}. 
	
	\subsection{Surrogate performance evaluation}\label{sec:performance_eva}
	
	A trained surrogate can be used to efficiently explore the behaviour of the ABM over the entire parameter space.	Relevant parameter combinations can then be selected for evaluation using the original ABM. Given the desire to avoid evaluating the computationally expensive true ABM, while also identifying positive calibrations, it is critical to maximize the performance of the surrogate to predict these calibrations. Recall that positive calibrations are points in the parameter space that fulfil the specific conditions, specified by an ABM modeller/user.  Such conditions might include any test that compares simulated output with real data (e.g. distance between real and simulated moments, a non-parametric test on distribution equality, mean squared prediction errors, etc.) and/or any specific feature the model might generate (e.g. fat tails in a specific distribution, growth rates of any variable above or below a given threshold, correlation patterns among a set of variables, etc.). In the two exercises presented in this paper (cf. Sections \ref{sec:BH} and \ref{sec:IS} below), both types of conditions are evaluated. 
	
An effective surrogate should maximize the ``True Positive Rate'' (TPR). Given a set of parameter combinations, the TPR measures the number of positive calibrations predicted by a learned surrogate model against the actual number of positive calibrations possible in the parameter space. 
	Automated hyper-parameter optimization procedures maximize the performance of the machine learning surrogate according to a learning score or metric.\footnote{Several procedures exist for tuning machine learning hyper-parameters, see e.g. \cite{feurer2015efficient}.} 
	Though our aim is to maximize the TPR of our surrogate, the scores used to train the surrogate depend on the particular form of the output condition. According to the two settings introduced above we distinguish between:

\begin{itemize}
\item \textbf{Binary outcome}. In this case the output of the calibration condition is discrete, such as Accept/Reject, and a measure of classification ability is needed. Specifically, we aim at maximizing the $F_1$-score.\footnote{Note that there is ``no free lunch'' with regard to performance measures, so their choice depends on the problem setting \citep[see e.g.][]{wolpert2002supervised} For a detailed description of the $F_1$-Score, see e.g. \cite{Van79}.} The $F_1$-score is an harmonic mean between $p$, which indicates the ratio between true positives and total positives and $r$, which represents the ratio of true positives to predicted ones: 	
\begin{equation}\label{eq:F1}
	F_1= 2\frac{p\cdot r}{p+r},
	\end{equation}
	The $F_1$-score takes a value between $0$ and $1$. In terms of Type I and Type II errors, it equates to:
	\begin{equation}\label{eq:F1_2}
	F_1= \frac{2\cdot\text{true positives}}{2\cdot \text{true positives} + \text{false positives} + \text{false negatives}}.
	\end{equation}
	
\item \textbf{Real-valued outcome}. In this case, our aim is to minimize the mean-squared error (MSE),
	\begin{equation}
	MSE=\frac{\sum_{i=1}^N(\hat{y}_i-y_i)^2}{N},
	\end{equation}
	where the surrogate predicts $\hat{y}_i$ over $N$ evaluation points with a true labelling $y$. We notice that this approach is in line, for instance, with \cite{Rec15}.
	
\end{itemize}	
	
	\subsection{Parameter importance}\label{section:param_imp}
The XGBoost algorithm employed in our surrogate modelling procedure allow us also to perform parameter sensitivity analysis at no costs. In particular, the machine learning algorithm provides an intuitive procedure of assessing the explained variance of the surrogate according to the relative number of times a parameter was ``split-on'' in the ensemble \citep[for details see e.g.][] {archer2008empirical,louppe2013understanding,breiman2001random}. As each tree is constructed according to an optimized splitting of the possible values for a specific parameter vector, and it is increasingly focusing on difficult-to-predict samples, splits dictate the relative importance of parameters in discriminating the output conditions of the ABM. Accordingly, the relative number of splits over a specific parameter provides a quantitative assessment of the surrogate model's sensitivity to the user-specified conditions specified by the parameter. This also allows a ranking over parameters on the basis of their relative importance in producing model behaviour that satisfies whatever conditions specified by the user. As this procedure is non-parametric, the resulting values should be interpreted as a rank-based statistic. In particular, the relative importance values associated to the number of splits only characterize the specific instantiation of the ensemble. The resulting counts provide insight into the relative performance for each parameter. A changing number of trees would result in a different number of splits for each parameter. As the number of trees approach infinity, the number of splits will converge to the true ratio of splits per parameter by the law of large numbers.

	\subsection{Training procedure}\label{section:training}
	
	The primary constraint we face is the limited number of parameter combinations that can be used for model evaluation (budget) without incurring in excessive computational costs. To address this issue, we propose a \textit{budgeted online active semi-supervised learning} approach that iteratively builds a training set of parameter vectors on which the agent-based model is actually evaluated in order to provide labelled data points for the training of the surrogate.  The aim of actively sampling the parameter space is to reduce the discrepancy between the regions that contain a manifold of interest and the function approximation produced by the surrogate model.  This semi-supervised learning approach \citep[see e.g.][]{zhu2005semi,goldberg2011oasis} minimizes the number of required evaluations, while improving the performance of the surrogate. Given that evaluated parametrizations are aggregated over several rounds and the stationary nature of the parameter space labels, we can use the $\log$ convergence results proved in \cite{ross2011reduction} to provide a guideline on the number of parametrizations to evaluate in each round. In particular, we evaluate $C\log{\text{budget}}$ parameters per round, with $C=1$, also ensuring at least one positive calibration in the initial seed round. Noting that the constant $C$ can be increased or decreased according to the particular ABM.
	
Generally, positive calibrations represent a very small percentage of points in the parameter space. For example, the concentration of positive calibrations in both of the ABMs presented in this paper represents less than $1\%$ of the parameters. Our approach exploits this imbalance by iteratively selecting a random subset of positive predicted calibrations over a finite number of rounds. As we use positive \textit{predicted} calibrations, we exploit semi-supervised learning with the surrogate to select which parameters should be evaluated in the next round. In order to maximize computing speed, the algorithm is initialized with a fixed subset of evaluated parameter combinations that are drawn according to a quasi-random Sobol sampling over the parameter space \citep{morokoff1994quasi}.\footnote{Note that in high dimensional spaces, standard design of experiments are computationally costly and show little or no advantage over random sampling \citep{bergstra2012random,lee15}.} Further, the number of samples are drawn according to the ``total variation'' analysis presented in \cite{saltelli2010variance}. These initial ``training'' points are then evaluated through the ABM, their labels recorded and finally used to initialize the first surrogate model. Once the surrogate is trained, new parameter combinations are sampled over the entire parameter space and labelled using the surrogate. A random subset of points $x_i$ are then selected from the predicted positive calibrations of the surrogate and evaluated for their true labels $y_i$ using the ABM. Given the $\log$ convergence rates presented in \cite{ross2011reduction}.

These new points are then added to the training set to train a new surrogate in the next round. This ``self-training'' procedure exploits the imbalance in the data to incrementally increase true positives, while reducing false positives. Note that this simple self-training procedure may result in no new predicted positives. In this case, the algorithm selects new points according to their predicted binary label entropy, where the latter is defined as the entropy between the predicted positive and negative calibration label probabilities. This incremental procedure continues until the targeted training budget is achieved. The algorithm pseudo-code is presented in Figure \ref{fig:pseudocode}.
	
	\begin{figure}[tbp]
		\begin{center}
				\fbox{
					\small
					\centering
					\begin{minipage}{5in}
						\small
						{\bfseries Set}: 
						\begin{itemize}
							\item Agent Based Model $\mbox{ABM}\in\mathbb{R}^J$
							\item Sampling distribution $\nu\in\mathcal{R}^J$
							\item Calibration function $C(\cdot)$
							\item Learning algorithm $\mathcal{A}$, with parameters $\Theta$
							\item Evaluation budget $B$
							\item Initial training set size $N\ll B$
							\item $X^{Training}\in\mathbb{R}^{N\times J}$
							\item Calibration labels $Y^{Training}\in\mathbb{N}^{N}$ binary outcome case (at least $1$ positive calibration)
							\item Calibration labels $Y^{Training}\in\mathbb{R}^{N}$ real-valued outcome case (at least $1$ positive calibration)
							\item Hyper-parameter optimization algorithm ($\text{HPO}$)
						\end{itemize}
						{\bfseries Initialize}:
						\begin{itemize}
							\item Per-round sampling size $S\ll B$										\item Per-round out-of-sample size $K\gg B$
						\end{itemize}
						{\bfseries While $|Y|<B$, repeat}
						\begin{enumerate}
							\item $\Theta = \text{HPO}(\mathcal{A}(\Theta,X^{Training},Y^{Training}))$
							\item Draw out-of-sample points $X^{OOS}\in\mathbb{R}^{K\times J}\sim\nu$
							\item Select $X^{sample}\in\mathbb{R}^{S\times J}$ from $X^{OOS}$																											\item Evaluate $X^{Training} = X^{Training} \cup X^{sample}$
							\item Evaluate $Y^{sample} = \{C(\mbox{ABM}(X^{sample}_i))\}_{i = 1\dots S}$
							\item Evaluate $Y^{Training} = Y^{Training} \cup Y^{sample}$
						\end{enumerate}
						{\bfseries end while}
					\end{minipage}
				}
			\end{center}
			\caption{Pseudo-code of our training algorithm. Note: $Y$ indicates labels; $X$ indicates parameter vectors. HPO: hyper parameter optimization; OOS: out of sample}
		\label{fig:pseudocode}
	\end{figure}
	
	%%%%%%%%%%%%%%%%%%%%%%%%%%%%%%%%%%%%%%%%%%%%%%%%%%%%%%%%
	\section{Surrogate modelling examples: The Brock and Hommes model}\label{sec:BH}
	
	In their seminal contribution, \citet{BH98} develop an asset pricing model (referred here as B\&H), where an heterogeneous population of agents trade generic assets according to different strategies (fundamentalist, chartists, etc.). We briefly introduce the model in (cf. Section \ref{subsec:BH_model}). Then we report the empirical setting (see Section \ref{subsec:empirical_BH}) and results of our machine learning calibration and exploration exercise (cf. Section \ref{subsec:BH_results}). Recall that the seed of the pseudo-random number generator is fixed and kept constant across runs of the model over different parameter vectors.	
	
	\subsection{The B\&H asset pricing model}\label{subsec:BH_model}
	
	There is a population of $N$ traders that can invest either in a risk free asset, which is perfectly elastically
	supplied at a gross return $R = (1 + r) > 1$, or in a risky one, which pays an uncertain dividend $y$ and has a price denoted by $p$. Wealth dynamics is given by
	\begin{equation}\label{eq:BH_wealth}
	W_{t+1}=RW_{t} + (p_{t+1} + y_{t+1} - Rp_t)z_{t},
	\end{equation}
	where $p_{t+1}$ and $y_{t+1}$ are random variables and $z_t$ is the number of the shares of the risky asset bought at time $t$.
	
	Traders are heterogeneous in terms of their expectations about future prices and dividends and are assumed to be myopic mean-variance maximizers. However, as information about past prices and dividends is publicly available in the market, agents can apply conditional expected value $E_t$, and variance $V_t$. The demand for share $z_{h,t}$ of agents with expectations of type $h$ is computed, solving:
	\begin{equation}\label{eq:BH_max}
	\max_{z_{h,t}} \left\{E_{h,t}(W_{t+1})-\frac{\nu}{2}V_{h,t}(W_{t+1})\right\},
	\end{equation}
	which in turns implies 
	\begin{equation}
	z_{h,t}=E_{h,t}(p_{t+1}+y_{t+1}-Rp_{t})/(\nu\sigma^2),
	\end{equation}
	where $\nu$ controls for agents' risk aversion and $\sigma$ indicates the conditional volatility, assumed to be equal across traders and constant over time.
	In case of zero supply of outside shares and different trader types, the market equilibrium equation can be written as:
	\begin{equation}\label{eq:BH_eq}
	Rp_t=\sum n_{h,t}E_{h,t}(p_{t+1}+y_{t+1}),
	\end{equation}
	where $n_{h,t}$ denotes the share that traders of type $h$ hold at time $t$.
	In presence of homogeneous traders, perfect information and rational expectations, one can derive the  no-arbitrage market equilibrium condition:
	\begin{equation}\label{eq:BH_arbitr}
	Rp^*_{t}=E_{t}(p^*_{t+1}+y_{t+1}) ,
	\end{equation}
	where the expectation is conditional on all histories of prices and dividends up to time $t$ and where $p^*$ indicates the fundamental price. Dividends are independent and identically distributed over time with constant mean, equation \eqref{eq:BH_arbitr} has a unique solution where the fundamental price is constant and equal to $p^*=E(y_t)/(R-1)$. In what follows, we will express prices as deviations from the fundamental price, i.e. $x_t=p_t-p^*_t$.
	
	At the beginning of each trading period $t = \{1,2, ..., T\}$, agents form expectations about future prices and dividends. Agents are heterogeneous in their forecasts. More specifically, investors believe that, in a heterogeneous world, prices may deviate from the fundamental value by some function $f_h(\cdot)$ depending upon past deviations from the fundamental price. Accordingly, the beliefs about $p_{t+1}$ and $y_{t+1}$ of agents of type $h$ evolve according to:
	\begin{equation}\label{eq:BH_forecast}
	E_{h,t}(p_{t+1}+y_{t+1})=E_t(p_{t+1}^*)+f_h(x_{t-1},...,x_{t-L}).
	\end{equation}
	Many forecasting strategies specifying different trading behaviours and attitudes  have been studied in the economic literature, \citep[see e.g.][]{Ban92, BH97, Lux00, Chi09}. \citet{BH98} adopt a simple linear representation of beliefs:
	\begin{equation}\label{eq:BH_beliefs}
	f_{h,t}=g_hx_{t-1} + b_h,
	\end{equation}
	where $g_h$ is the trend component and $b_h$ the bias of trader type $h$. If $b_h\neq 0$, the agent $h$ can be either a pure trend chaser if $g_h>0 $ (strong trend chaser if $g>R$), or a contrarian if
	$g<0$ (strong contrarian if $g<R$). If $g_h\neq 0$, the agent of type $h$ is purely biased
	(upward or downward biased if $b_h>0$ or $b_h<0)$.  In the special case when both $g_h$ and $b_h$ are equal to zero, the agent is a ``fundamentalists'', i.e. she believes that prices return to their fundamental value. Agents can also be fully rational, with $f_{rational,t}=x_{t+1}$. In such a case, they have perfect foresight but, they must pay a cost $C$.\footnote{In our experiments we allow for the possibility that a positive cost might be by paid also by non-rational traders. This mirrors the fact that some trader might want to buy additional information, which they might not be able to use (due e.g. to computational mistakes).}
	
	In our application, we use a simple model with only two types of agents, whose behaviours vary according to the choice of trend components, biases and perfect forecasting costs.
	Combining equations \eqref{eq:BH_eq}, \eqref{eq:BH_forecast} and \eqref{eq:BH_beliefs}, one can derive the following equilibrium condition:
	\begin{equation}\label{eq:BH_equilbrium_final}
	Rx_t=n_{1,t}f_{1,t}+n_{2,t}f_{2,t},
	\end{equation}
	which allows to compute the price of the risky asset (in deviation from the fundamental) at time $t$.
	
	Traders switch among different strategies according to the their evolving profitability.  More specifically, each strategy $h$ is associated with a fitness measure of the form:
	\begin{equation}\label{eq:BH_fitness}
	U_{h,t}=(p_t+y_t-Rp_{t-1})z_{h,t}-C_h+\omega U_{h,t-1}
	\end{equation}
	where $\omega\in[0,1]$ is a weight attributed to past profits.  At the beginning of each period, agents reassess the profitability of their trading strategy with respect to the others. The probability that an agent choose strategy $h$ is given by:
	\begin{equation}\label{eq:BH_shares}
	n_{h,t}=\frac{\exp(\beta U_{h,t})}{\sum_h\exp(\beta U_{h,t)}},
	\end{equation}
	where the parameter $\beta \in [0,+\infty)$ captures traders' intensity of choice. According to equation \ref{eq:BH_shares}, successful strategies gain an increasing number of followers. In addition, the algorithm introduces a certain amount of randomness,
	as less profitable strategies may still be chosen by traders. In this way, the model captures imperfect information and agents' bounded rationality. Moreover, the system can never be stacked in an equilibrium where all traders adopt the same strategy.

	\subsection{Experimental design and empirical setting}\label{subsec:empirical_BH}
	Despite the model is relatively simple, different contributions have tried to match the statistical properties of its output with those observed in real financial markets \citep{Bos07, Rec15, Lamperti16, Kuk16}. This makes the model an ideal test case for our surrogate: it is relatively cheap in terms of computational needs, it offers a reasonably large parameter space and it has been extensively studied  in the literature.
	
	There are 12 free parameters (Table \ref{tab:param_BH}) determined through calibration.\footnote{We underline that the dimension of the parameter space is in line or even larger that in recent studies on ABM meta-modelling \citep[see e.g.][]{Salle14,Bar16}.} The ranges for parameters' values have been identified relying on both economic reasoning and previous experiments on the model. However, their selection is ultimately a user specific decision.  Our procedure manages the computational constraints faced by modellers working with large parameter spaces. In what follows, we refer to the parameter space spanned by the intervals specified in the last column of Table \ref{tab:param_BH}. Naturally, it can be further expanded or reduced according to the user's needs and the available budget.
	
	\begin{table}[tbp]
		\caption{Parameters and explored ranges in the Brock and Hommes model.}\label{tab:param_BH}
		\centering
		\adjustbox{max width=\textwidth, max height=\textheight}{
			\begin{tabular}{clcc}
				\toprule
				\textbf{Parameter} & \textbf{Brief description} & \textbf{Theoretical support} & \textbf{Explored range}  \\\midrule\midrule
				\multicolumn{4}{c}{Brock and Hommes Model}\\
				\midrule
				$\beta$  & intensity of choice &  $[0;+\infty)$  & $[0.0; 10.0]$ \\
				$n_1$    & initial share of type 1 traders& $[0; 1]$ & 0.5\\
				$b_1$    & bias of type 1 traders & $(-\infty; +\infty)$ & $[-2.0;2.0]$ \\
				$b_2$    & bias of type 2 traders &  $(-\infty; +\infty)$ & $[-2.0;2.0]$\\
				$g_1$    & trend component of type 1 traders & $(-\infty; +\infty)$ & $[-2.0;2.0]$\\
				$g_2$    & trend component of type 2 traders &$(-\infty; +\infty)$ & $[-2.0;2.0]$\\
				C     & cost of obtaining type 1 forecasts &$[0;+\infty)$ & $[0.0; 5.0]$\\
				$\omega$     & weight to past profits &  $[0.0, 1.0]$  & $ [0.0; 1.0]$ \\
				$\sigma$ & asset volatility &  $(0; +\infty)$ & $(0.0; 1.0]$\\
				$\nu$ & attitude towards risk & $[0; +\infty]$ & $[0; 100]$\\
				r     & risk-free return &  $(1; +\infty)$  &  $[1.01, 1.1]$ \\
				T$_{BH}$ & number of periods & $\mathcal{N}$  & 500\\
				\bottomrule
			\end{tabular}
		}
	\end{table}
	
 Let us now consider the conditions identifying positive calibrations. As already discussed above, any feature of model's output can be employed to express such conditions. According to Section \ref{sec:surrogate} two types of calibration criteria are considered, giving respectively binary and real-valued outcomes. In the binary outcome case, we employ a two samples Kolmogorov-Smirnov (KS) test between the distribution of logarithmic returns obtained from the numerical simulation of the model and the one obtained from real stock market data.\footnote{Let $p_t$ and $p_{t-1}$ be the prices of an asset at two subsequent time steps. The logarithmic return from $t-1$ to $t$ is given by $r_t=\log(p_t/p_{t-1})\simeq (p_t-p_{t-1})/p_{t-1}$.} More specifically, we rely on daily adjusted closing prices for the S\&P 500 from December 09, 2013 to December 07, 2015, for a total of 502 observations, and we compute the following test statistic:\footnote{The data have been obtained from Yahoo Finance: \url{https://finance.yahoo.com/quote/\%5EGSPC/history}. The test is passed if the null hypothesis ``equality of the distributions'' is not rejected at 5\% confidence level.}
	\begin{equation}\label{eq:KS_test}
	D_{{RW},{S}}=\sup_r |F_{RW}(r) - F_{S}(r)|,
	\end{equation}
	where $r$ indicates logarithmic returns and $F_{RW}$ and $F_{S}$ are the empirical distribution functions of the real world ($RW$) and simulated ($S$) samples respectively. Then, in a real-valued outcome setting, we use the p-value of the KS test, $P(D>D_{RW,S})$, as an expression of model's fit with the data. We also consider an equivalent condition for binary outcomes, where predicted labels with a p-value above $5\%$ are considered as positive calibrations. This choice is intentional as equivalent conditions allow a comparison between the binary and real-valued outcomes in terms of precision (ability to identify true calibrations) and computational time (in the real-valued scenario there is more information to be processed.)   
	
	We train the surrogate 100 times over 10 different budgets of
	250, 500, 750, 1000, 1250, 1500, 1750, 2000, 2250, 2500 labelled parameter combinations and evaluate it on 100000 unlabelled points. Having a large number of out-of-sample, unlabelled, possibly well-spread points is fundamental to evaluate the performance of the meta-model. We use a larger evaluation set than any other meta-modelling contribution we are aware of \citep[see, for instance,][]{Salle14, Dosi16, Bar16}.

	\subsection{Results}\label{subsec:BH_results}

\begin{figure}[tbp]
		\centering
		\includegraphics[scale=0.5]{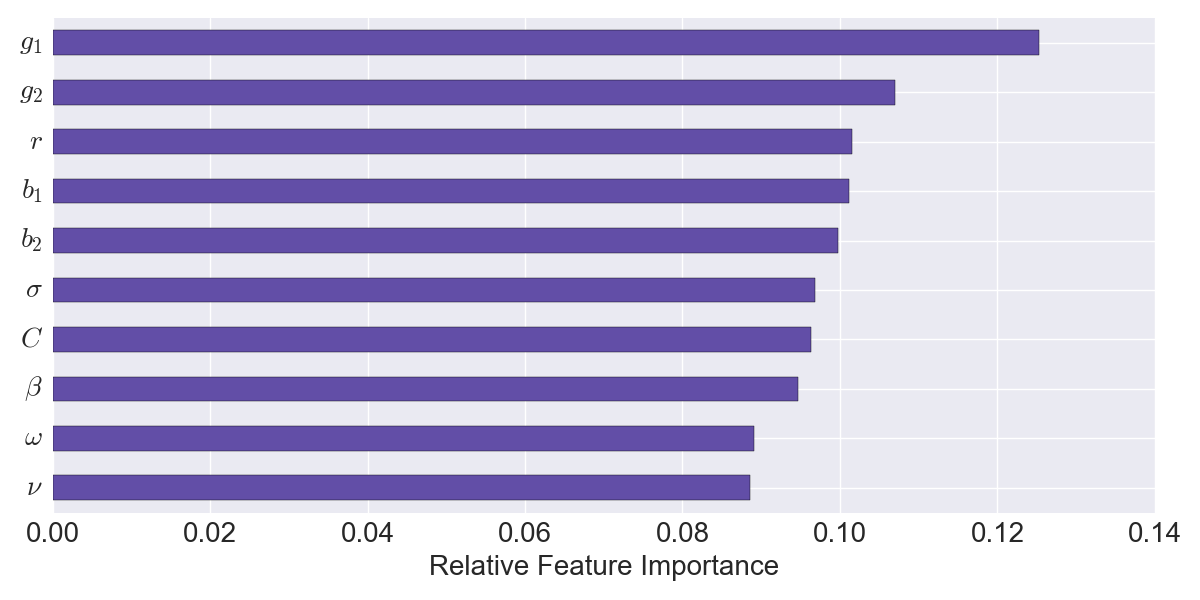}
		\caption{Importance of each parameter (feature) in shaping behaviour of the Brock and Hommes model according to the specified conditions (i.e. equality between distributions of simulated and real returns). As noted in Section \ref{section:param_imp}, this chart demonstrates the relative rank-based importance for each parameter.}
		\label{fig:BH_feat_impo}
	\end{figure}

	In Figure \ref{fig:BH_feat_impo}, we show the parameter importance results for the Brock and Hommes (B\&H) model. We find that the most relevant parameters to fit the empirical distribution of returns observed in the SP500 are those characterizing traders' attitude towards the trend ($g_1$ and $g_2$) and, secondly, their bias ($b_1$ and $b_2$). This result is in line with recent findings by \citet{Rec15} and \citet{Lamperti16} obtained using the same model. Moreover, the ``intensity of choice'' parameter ($\beta$, cf. Section \ref{sec:BH}) is of crucial importance in the original model developed by \cite{BH98}, but does not appear to be particularly relevant in determining the fit of the model with the data when compared to other behavioural parameters (at least within the range expressed by Table \ref{tab:param_BH})\footnote{See also \citet{Bos07} where the authors estimate the B\&H model on the SP500 and, in many exercises, find the switching parameter not to be significant.}. Also traders' risk attitude ($\alpha$) and the weight associated to past profits ($\omega$) are relatively unimportant to shape the empirical performance of the model. 
	
\begin{figure}
		\begin{subfigure}[b]{0.5\textwidth}
			\includegraphics[scale=0.3, trim={25mm 21mm 0mm 29mm}, clip]{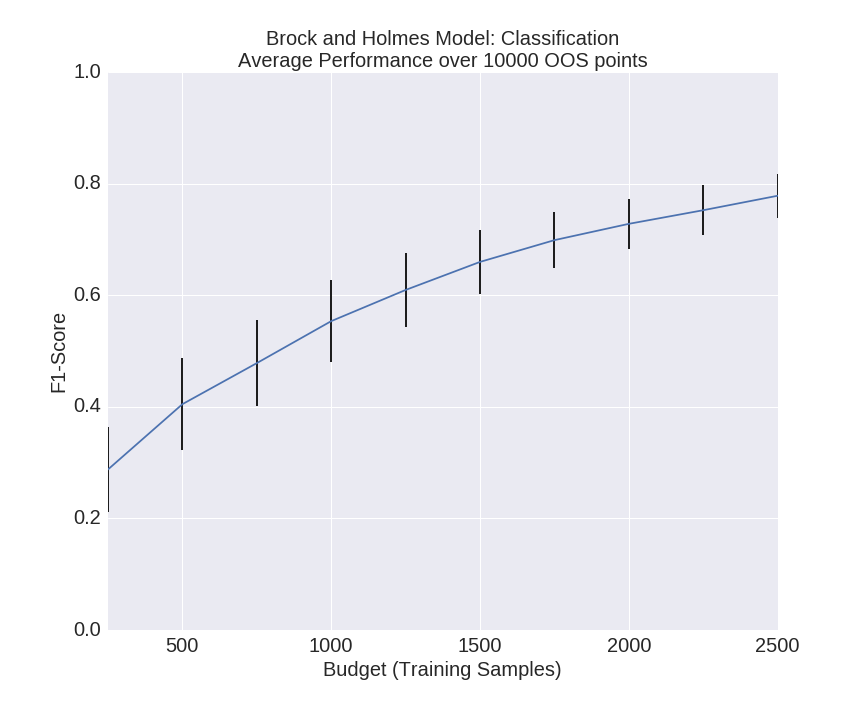}
			\caption{Binary-outcome: F1 Score\vspace{10pt}}\label{fig:BH_f1}
		\end{subfigure}%
		\begin{subfigure}[b]{0.5\textwidth}
			\includegraphics[scale=0.3, trim={17mm 21mm 0mm 29mm}, clip]{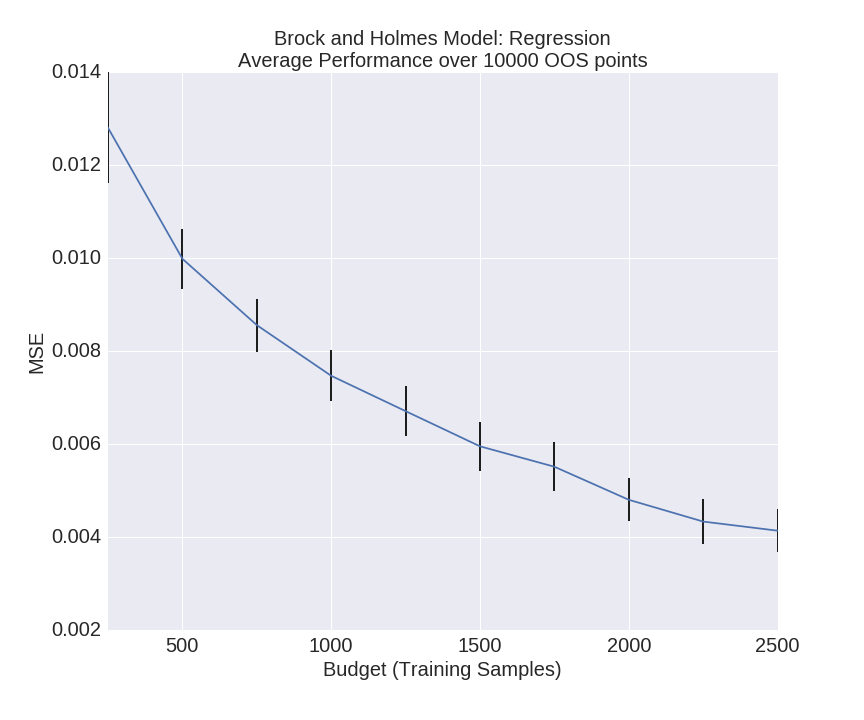} 
			\caption{Real-valued outcome: Mean Squared Error\vspace{10pt}}\label{fig:BH_MSE}
		\end{subfigure}\\
		\begin{subfigure}[b]{0.5\textwidth}
			\includegraphics[scale=0.3, trim={25mm 21mm 0mm 28mm}, clip]{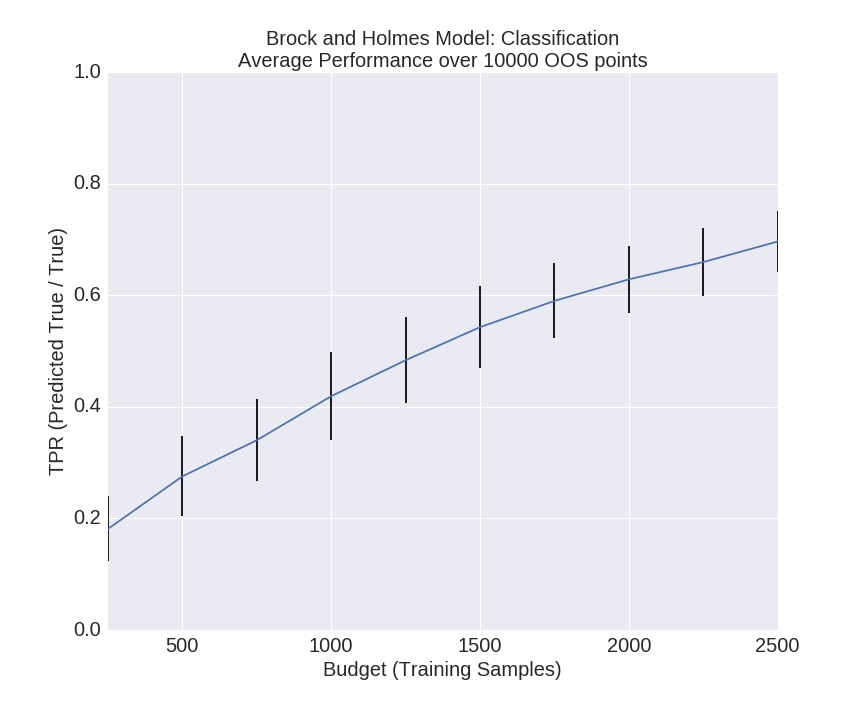}
			\caption{Binary-outcome: True Positive Rate\vspace{10pt}}\label{fig:BH_TPR_class}
		\end{subfigure}%
		\begin{subfigure}[b]{0.5\textwidth}
			\includegraphics[scale=0.3, trim={25mm 21mm 0mm 28mm}, clip]{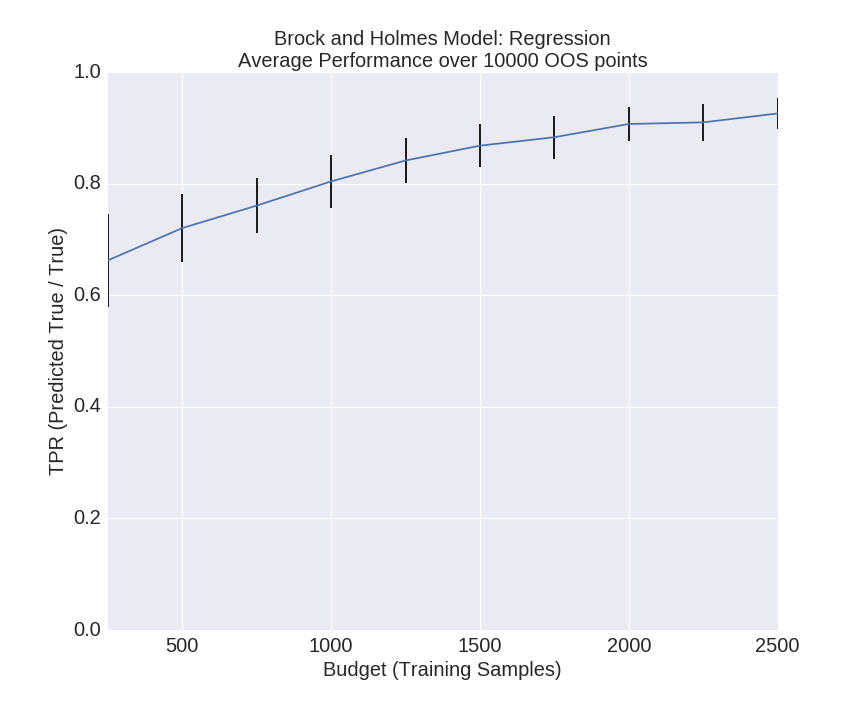}
			\caption{Real-valued outcome: True Positive Rate\vspace{10pt}}\label{fig:BH_TPR_reg}
		\end{subfigure}\\
		\begin{subfigure}[b]{0.5\textwidth}
			\includegraphics[scale=0.3, trim={26mm 21mm 0mm 28mm}, clip]{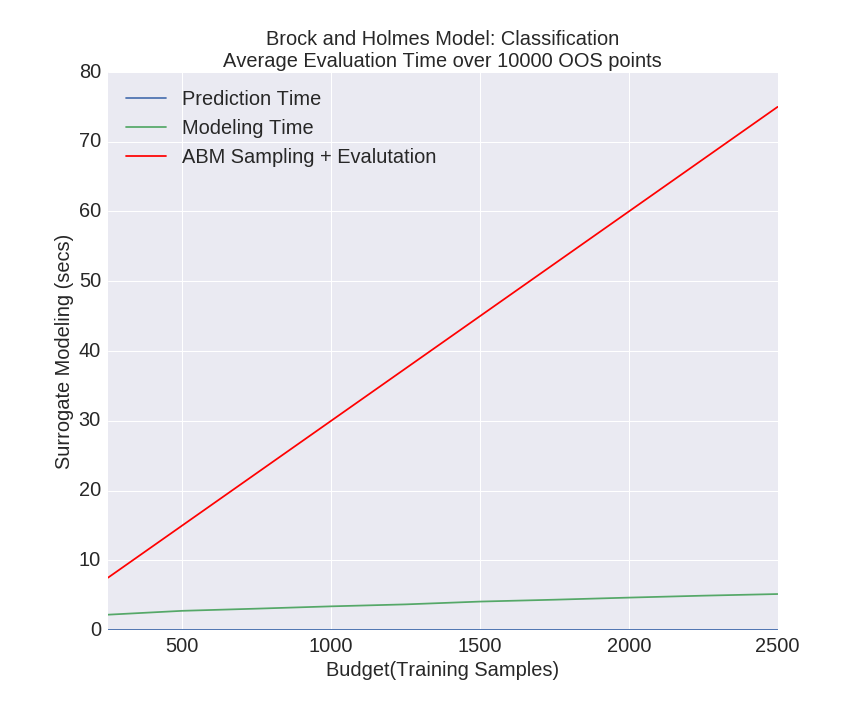}
			\caption{Binary-outcome: Computation Time\vspace{10pt}}\label{fig:BH_time_class}
		\end{subfigure}%
		\begin{subfigure}[b]{0.5\textwidth}
			\includegraphics[scale=0.3, trim={26mm 21mm 0mm 28mm}, clip]{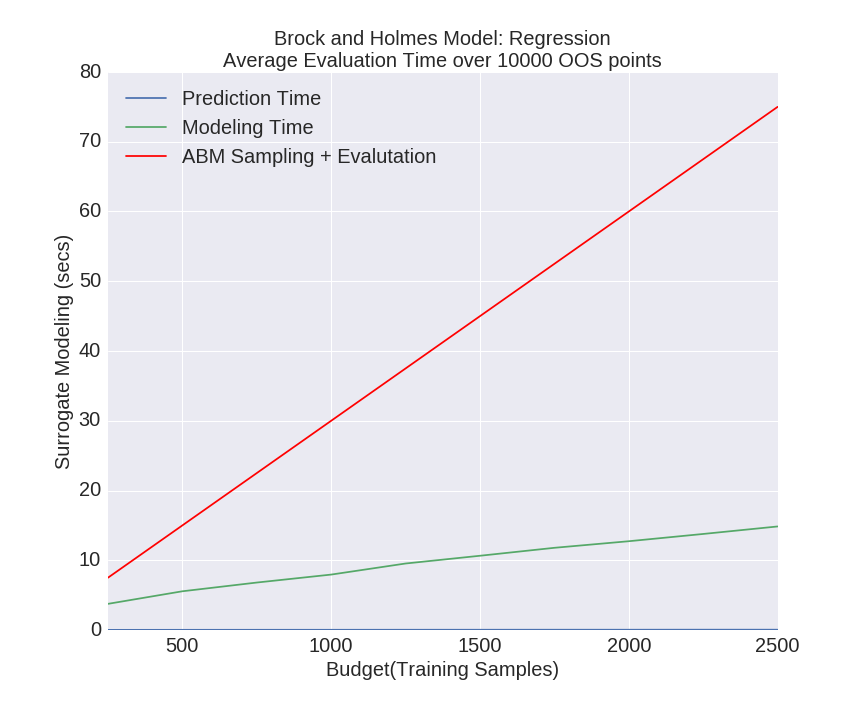}
			\caption{Real-valued outcome: Computation Time\vspace{10pt}}\label{fig:BH_time_reg}
		\end{subfigure}
		\caption{Brock and Hommes surrogate modelling performance averaged over a pool of 10000 parametrizations. Black vertical lines indicate 95\% confidence intervals on 100 repeated and independent experiments. } 
		\label{fig:results_BH}
	\end{figure}

	Let us now consider the behaviour of the surrogate. As outlined in Section \ref{sec:performance_eva}, we run a series of exercises where the surrogate is employed to explore the behaviour of the model over the parameter space and filter out positive calibrations matching the distribution of real stock-market returns. Figure \ref{fig:results_BH} collects the results and show the performance of the surrogate in the two proposed settings (binary and real-valued outcome). Within the binary outcome exercise, the $F_1$-score increases with the size of the training sample and reaches a peak of around 0.80 when 2500 points are employed (cf. Figure \ref{fig:BH_f1}). In other words, the average between the share of true positive calibrations and the share of positive calibrations the surrogate correctly predicts is 0.8. Taking into consideration the upper bound of 1 and various practical applications \citep[e.g.][]{petro, cire}, we consider the former result satisfying.  However, such a classification performance should be evaluated in view of the surrogate's \textit{searching ability}, which is reported in Figure \ref{fig:BH_TPR_class} and indicates the share of total positive calibration that the surrogate is able to find.  Specifically, we find that around 75\% of the positive calibrations present in the large set of out-of-sample points are found.  
	
	Obviously, the surrogate's performance worsens as the training sample size is reduced. However, once we move to the real-valued setting, where the surrogate is learned using a continuous variable (containing more information about model's behaviour), its performance is remarkably higher. Indeed, even when the sample size of the training points is particularly low (500), the True Positive Ratio (TPR) is around 70\%, and reaches almost 95\% (on average) when 2500 parameter vectors are employed (see Figure \ref{fig:BH_TPR_reg}). 
	
	Timing results are reported according to the average number of seconds required for a single compute core to complete the specific task 100 times. The increase in performance from classification (see Figure \ref{fig:BH_time_class}) to regression (see Figure \ref{fig:BH_time_reg}) requires roughly 3X the modelling time and a nearly equivalent prediction time. Given this negligible prediction time, our approach facilitates a nearly costless exploration of the parameter space, delivering good results in terms of $F_1$-score, TPR and MSE. The time savings in comparison to running the original ABM are substantial. In this exercise over a set of 10000 out-of-sample points, the surrogate is 500X faster on average in prediction. Note also that the learned surrogate is reusable on any number of out-of-sample parameter combinations, without the need for additional training. Further, we remark that computational gains are expected to be larger as more complex and expensive-to-simulate models are used. The next section goes in this direction.  
	
	\section{Surrogate modelling examples: the Islands model }\label{sec:IS}
	
	In the ``Island'' growth model \citep{FD03}, a population of heterogeneous firms
	locally interact discovering and diffusing new technologies, which ultimately lead to the emergence (or not) of endogenous growth. After presenting the model in Section \ref{section:island_model}, we describe the empirical setting (see Section \ref{subsec:empirical_IS}) and the results of the machine learning calibration and exploration exercises (cf. Section \ref{Section:island_results}). Recall that the seed used for the pseudo-random number generator is fixed and kept constant across runs of the model over different parameter vectors.
	
	\subsection{The Island growth model}\label{section:island_model}
	
	A fixed population of heterogeneous firms ($I=1, 2,..., N$) explore an unknown technological space (``the sea''), punctuated by islands (indexed by $j=1, 2,...$) representing new technologies. The technological space is represented by a 2-dimensional, infinite, regular lattice endowed with the Manhattan metrics $d_1$. The probability that each node $(x, y)$ is an island is equal to $p(x, y)=\pi$. There is only one homogeneous good, which can be ``mined'' from any island.  Each island is characterized by a productivity coefficient $s_j=s(x,y)>0$. The production of agent $i$ on island $j$ having coordinates $(x_j,y_j)$ is equal to: 
	\begin{equation}
	\label{eq:island_production}
	Q_{i,t}=s(x_j,y_j)[m_t(x_j,Y_j)]^{\alpha-1},
	\end{equation} 
	where $\alpha\geq1$ and $m_t(x_j,y_j)$ indicates the total number of miners working on $j$ at time $t$. The GDP of the economy is simply obtained by summing up the production of each island. 
	
	Each agent can choose to be a \textit{miner} and produce an homogeneous final good in her current island, to become an \textit{explorer} and search for new islands (i.e. technologies), or to be an \textit{imitator} and moves towards a known island. In each time step, miners can decide to become explorer with probability $\epsilon>0$. In that case, the agent leaves the island and ``sails'' around until another (possibly still unknown) island is discovered. During the search, explorers are not able to extract any output and randomly move in the lattice. When a new island (technology) is discover, its productivity is given by:
	\begin{equation}\label{eq:island_productivity}
	s_{j^{ \text{new} }}=(1+W)\{[|x_{j^{ \text{new} }}| + |y_{j^{ \text{new} }}|]+\varphi Q_i+\omega\}
	\end{equation}
	where $W$ is a Poisson distributed random variable with mean $\lambda>0$, $\omega$ is a uniformly distributed random variable with zero mean and unitary variance, $\varphi$ is a constant between zero and one and, finally, $Q_i$ is the output memory of agent $i$. Therefore, the initial productivity of a newly discovered island depends on four factors \citep[see][]{Dos88}: (i) its distance from the origin; (ii) cumulative learning effects ($\phi$); (iii) a random variable $W$ capturing radical innovations (i.e. changes in technological paradigms); (iv) a stochastic i.i.d. zero-mean noise controlling for high-probability low-jumps (i.e. incremental innovations).
	
	Miners can also decide to imitate currently available technologies by taking
	advantage of informational spill-overs stemming from more productive islands located
	in their technological neighbourhoods. More specifically, agents
	mining on any colonized island deliver a signal, which is instantaneously spread in
	the system. Other agents in the lattice receive the signal with probability:
	\begin{equation}
	\label{eq:islands_prob_signal}
	w_t(x_j,y_j ; x, y)=\frac{m_t(x_j, y_j)}{m_t}\exp \{-\rho[|x-x_j| + |y-y_j|]\},
	\end{equation} 
	which depends on the magnitude of technology gap as well as on the physical distance between two islands ($\rho>0$). Agent $i$ chooses the strongest signal and  become an imitator sealing to island according to the shortest possible path. Once the imitated island is reached, the imitator will start mining again.
	
	The model shows that the very possibility of notionally unlimited (albeit	unpredictable) technological opportunities is a necessary condition for the emergence of endogenous exponential growth. Indeed, self-sustained growth is achieved whenever technological opportunities (captured by both the density of islands $\pi$ and the likelihood of radical innovations $\lambda$), path-dependency (i.e. the fraction of idiosyncratic knowledge, $\varphi$, agents carry over to newly discovered technologies), and spreading intensity in the information diffusion process ($\rho$), are beyond some minimum thresholds \citep{FD03}. Moreover, the system endogenously generate exponential growth if the trade-off between exploration and exploitation is solved, i.e. if the ecology of agents find the right balance between searching for new technologies and mastering the available ones.

	\subsection{Experimental design and empirical setting}\label{subsec:empirical_IS}
	
	The Island model employs eight input parameters to generate a wide array of growth dynamics. We report the parameters, their theoretical support and the explored range in Table \ref{tab:param_IS}. We kept the number of firms fixed (and equal to 50) to study what happens to the same economic system, when the parameters linked to behavioural rules are changed.\footnote{Note that the Island model does not exhibit scale effects: the results generated by the model does not depend on the number of agents in the system \citep{FD03}.}
	
	\begin{table}[tbp]
		\caption{Parameters and explored ranges in the Island model.}\label{tab:param_IS}
		\centering
		\adjustbox{max width=\textwidth, max height=\textheight}{
			\begin{tabular}{clcc}
				\toprule
				\textbf{Parameter} & \textbf{Brief description} & \textbf{Theoretical support} & \textbf{Explored range}  \\\midrule\midrule
				\multicolumn{4}{c}{Islands Model}\\\midrule
				$\rho$  & degree of locality in the diffusion of knowledge & $[0, + \infty)$ & $[0; 10]$  \\
				
				$\lambda$ & mean of Poisson r.v. - jumps in technology & $[0; +\infty)$ & 1\\
				$\alpha$ & productivity of labour in extraction & $[0, +\infty)$ & $[0.8; 2]$ \\
				$\varphi$  & cumulative learning effect & $[0, 1]$ & $[0.0; 1.0]$ \\
				$\pi$   & probability of finding a new island &$ [0.0, 1.0]$ & $[0.0; 1.0]$ \\
				$\epsilon$ & willingness to explore & $[0, 1]$ & $[0.0; 1.0]$ \\
				$ m_0$    & initial number of agents in each island & $[2, +\infty)$ & 50 \\
				$ T_{IS}$     & number of periods & $\mathcal{N} $    & 1000\\
				\bottomrule
			\end{tabular}
		}
	\end{table}

	\begin{figure}[tbp]
		\centering
		\includegraphics[scale=0.5]{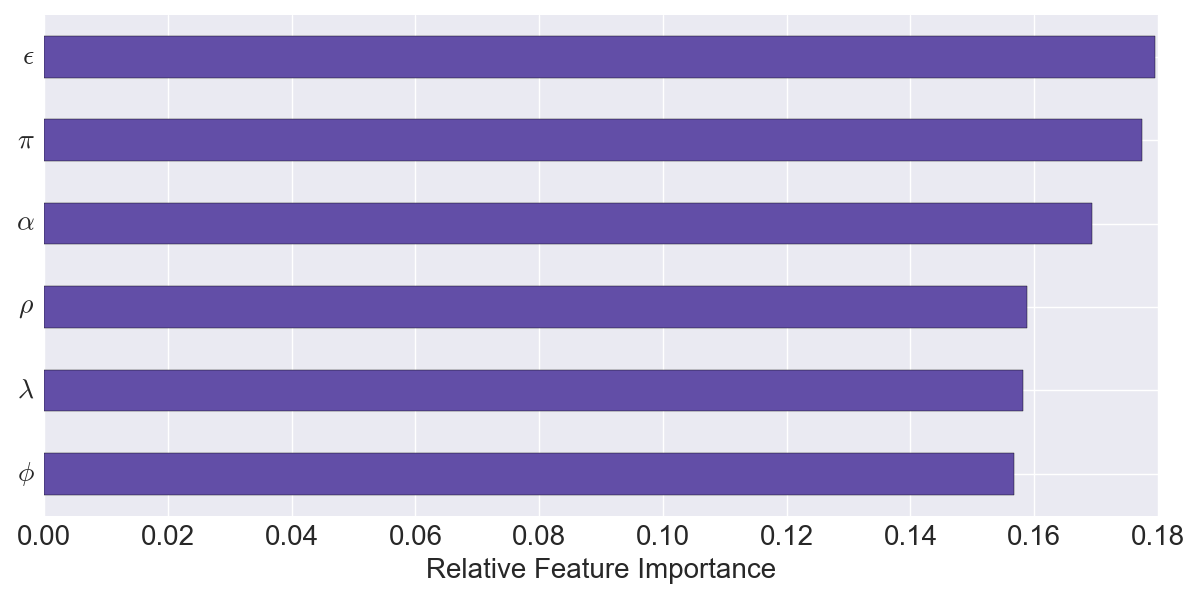}
		\caption{Importance of each parameter (feature) in shaping behaviour of the Islands model according to the specified conditions (sustained growth and fat tails). As noted in Section \ref{section:param_imp}, this chart demonstrates the relative rank-based importance for each parameter.}
		\label{fig:IS_feat_impo}
	\end{figure}

	\begin{figure}
		\begin{subfigure}[b]{0.5\textwidth}
			\includegraphics[scale=0.3, trim={25mm 22mm 0mm 28mm}, clip]{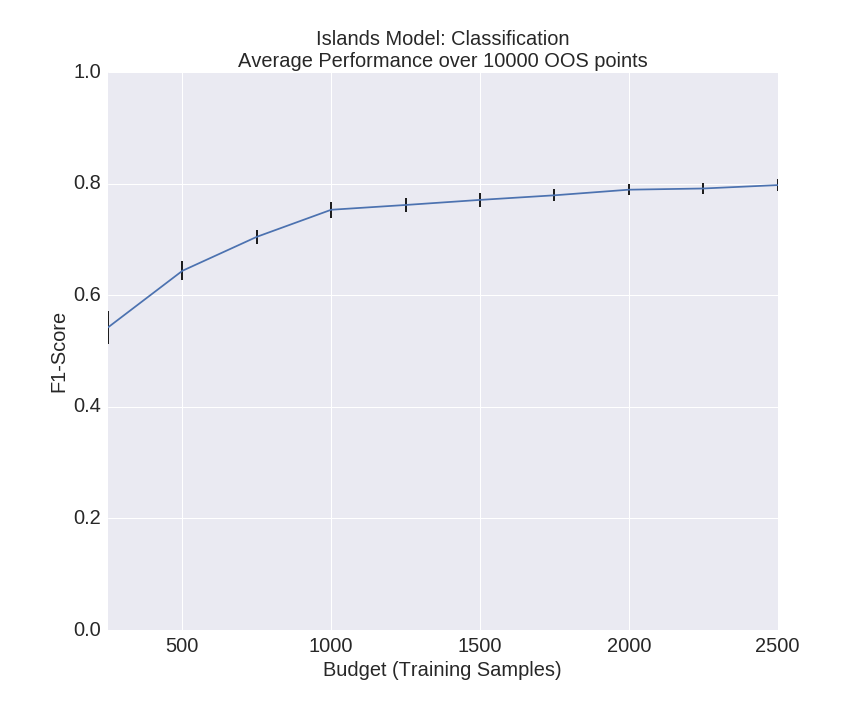}
			\caption{Binary-outcome: F1 Score\vspace{10pt}}\label{fig:IS_f1}
		\end{subfigure}%
		\begin{subfigure}[b]{0.5\textwidth}
			\includegraphics[scale=0.3, trim={20mm 22mm 0mm 28mm}, clip]{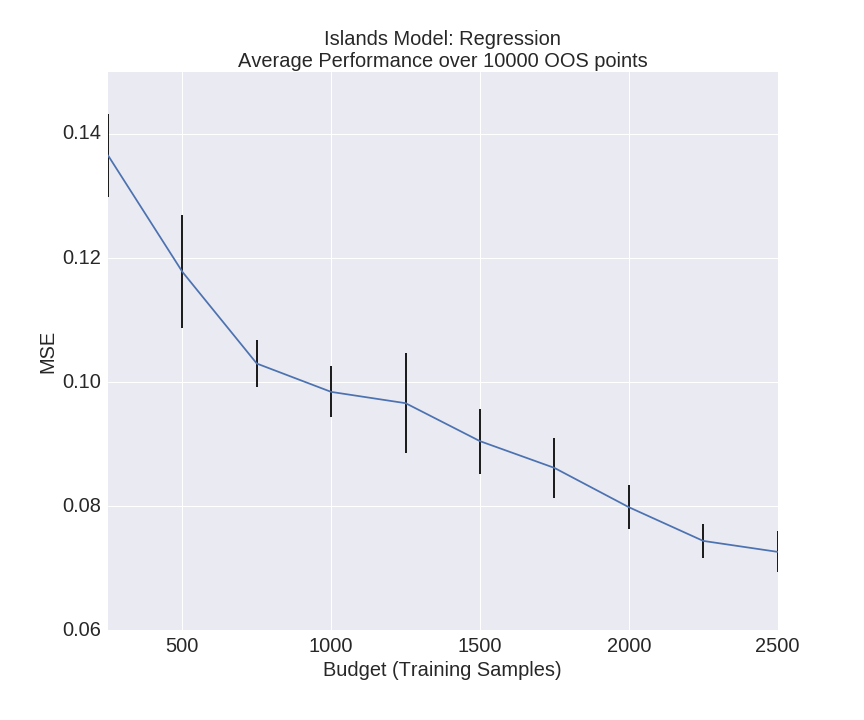} 
			\caption{Real-valued outcome: Mean Squared Error\vspace{10pt}}\label{fig:IS_MSE}
		\end{subfigure}\\
		\begin{subfigure}[b]{0.5\textwidth}
			\includegraphics[scale=0.3, trim={25mm 22mm 0mm 28mm}, clip]{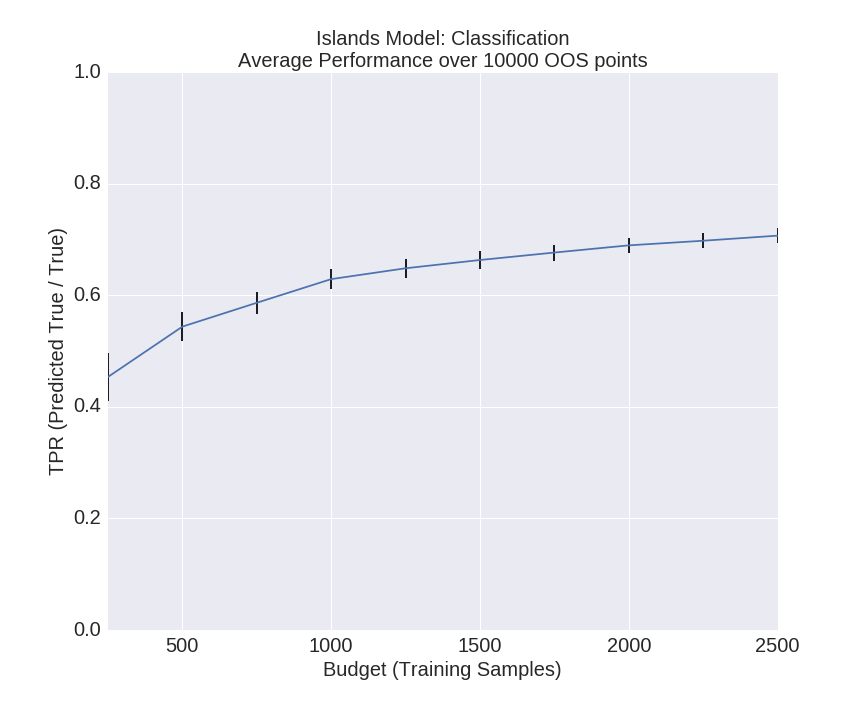}
			\caption{Binary-outcome: True Positive Rate\vspace{10pt}}\label{fig:IS_TPR_class}
		\end{subfigure}%
		\begin{subfigure}[b]{0.5\textwidth}
			\includegraphics[scale=0.3, trim={25mm 22mm 0mm 28mm}, clip]{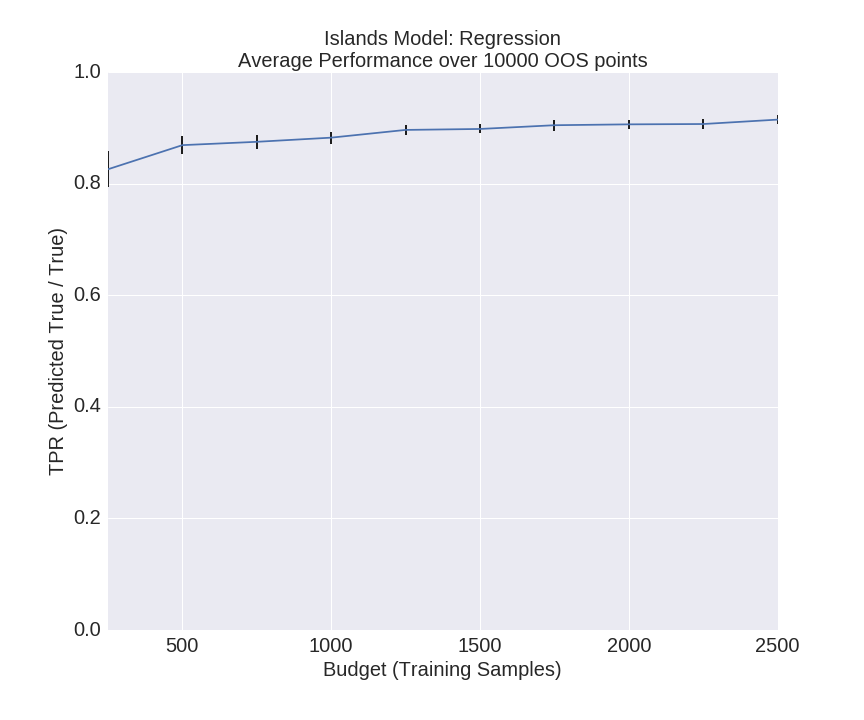}
			\caption{Real-valued outcome: True Positive Rate\vspace{10pt}}\label{fig:IS_TPR_reg}
		\end{subfigure}\\
		\begin{subfigure}[b]{0.5\textwidth}
			\includegraphics[scale=0.3, trim={23mm 22mm 0mm 28mm}, clip]{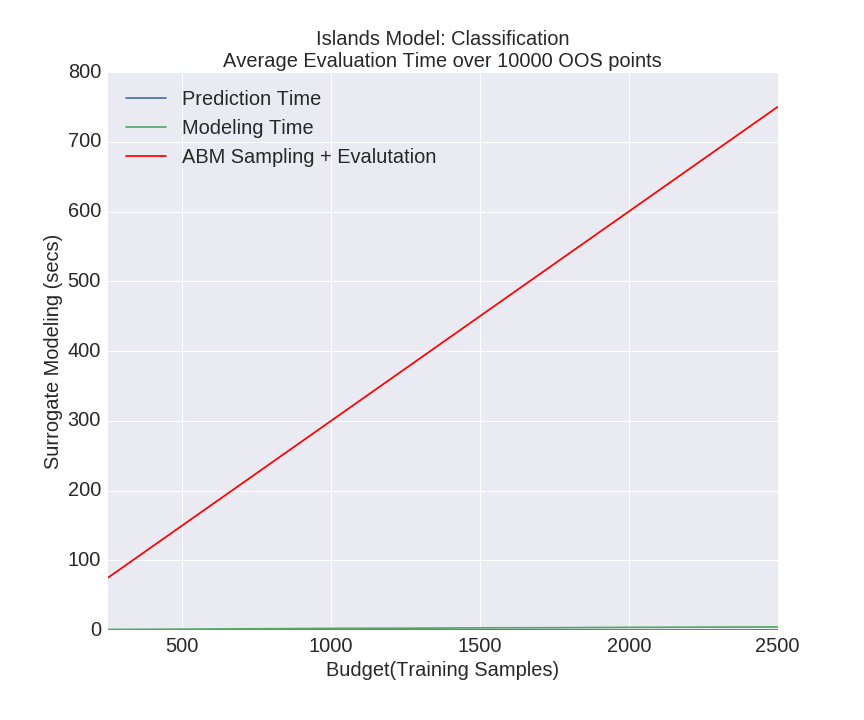}
			\caption{Binary-outcome: Computation Time\vspace{10pt}}\label{fig:IS_time_class}
		\end{subfigure}%
		\begin{subfigure}[b]{0.5\textwidth}
			\includegraphics[scale=0.3, trim={23mm 22mm 0mm 28mm}, clip]{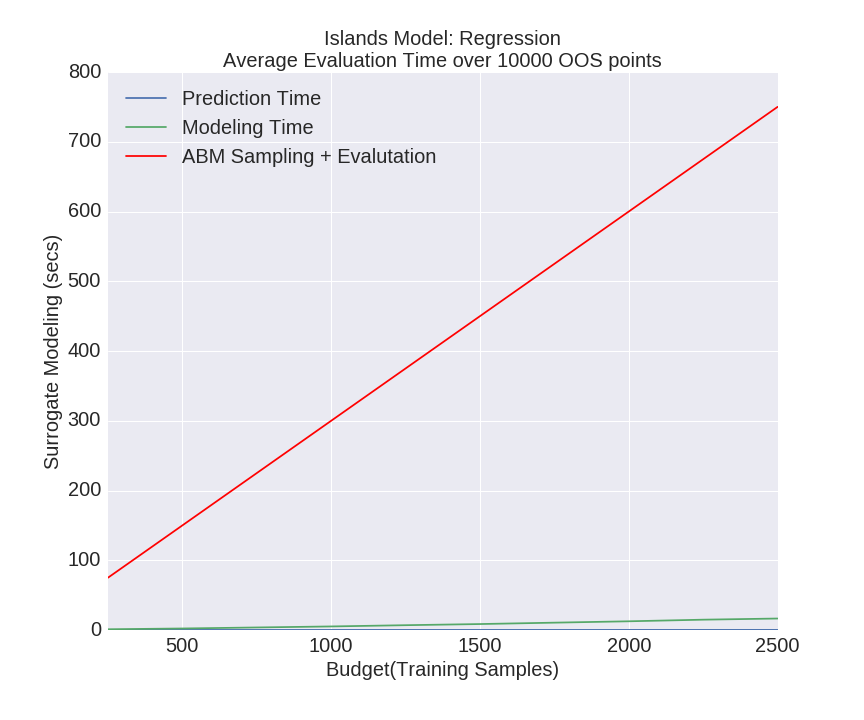}
            \caption{Real-valued outcome: Computation Time\vspace{10pt}}\label{fig:IS_time_reg}
		\end{subfigure}
		\caption{Islands surrogate modelling performance versus budget size averaged over a pool of 10000 parametrizations. Black vertical lines indicate 95\% confidence intervals on 100 repeated and independent experiments.} 
		\label{fig:results_IS}
	\end{figure}

	Similarly to section \ref{subsec:empirical_BH}, we characterize a binary outcome and a real-valued outcome setting. In the first case, the surrogate is learnt using a binary target variable $y$ taking value 1 if a user-defined specific set of conditions is satisfied and zero otherwise. More specifically, we define two conditions characterizing the GDP time series generated by the model. The first condition requires the model to generate self-sustained sustained pattern of output growth. Given the long-run average growth rate of the economy ($AGR$):
	\begin{equation}
	AGR=\frac{\log(GDP_T)-\log(GDP_1)}{T-1},
	\end{equation}
	sustained growth emerges if $AGR>2\%$.
	
	The second condition aims at capturing the presence of fat tails in the output growth-rate distributions. These empirical regularities suggest that deep downturns can coexist with mild fluctuations and has been found in both OECD \citep{FNR08} and developing countries \citep{CD09, LM16}. More specifically, we fit a symmetric exponential power distribution \citep[see][]{subbo,Bose} , whose functional form reads: 
	\begin{equation}\label{eq:subbo}
	f(x)=\frac{1}{2ab^{\frac{1}{b}}\Gamma(1+\frac{1}{b})}e^{-\frac{1}{b}|\frac{x-\mu}{a}|^{b}}
	\end{equation}
	where $a$ controls for the standard deviation, $b$ for the shape of the distribution and $\mu$ represents the mean. As $b$ gets smaller, the tails become fatter. In particular, when $b=2$ the distribution reduces to a Gaussian one, while for $b = 1$ the density is Laplacian. We say that the output growth-rate distribution exhibits fat tails if $b\leq 1$. Note that there is a hierarchy in the conditions we have just defined: only those parametrizations satisfying the first one ($AGR>2\%$) are retained as candidates for positive calibrations and further investigated with respect to the second condition. In the real-valued outcome case, instead, we just focus on shape of growth rates distribution. In particular, we our target variable is the estimated $b$ of the symmetric power exponential distribution and a positive calibration is found if $b>1$.\footnote{In the real-valued outcome setting our exercise is comparable to those performed in \citet{Dosi16}, where the same distribution and parameters are used in a model of industrial dynamics.} Again, the choice of the condition to be satisfied ensures (partial, in this case) consistency between the two settings.
	
	We train the surrogate as we did with the B\&H model, but given the higher computational complexity of the Island model, we reduce the number of unlabelled points to 10000.\footnote{This choice is motivated by the fact that  we need to run the model on the out-of-sample points in order to evaluate the surrogate.} 	
	
\subsection{Results}\label{Section:island_results}
	
	As for the Brock and Hommes model, we start our analysis reporting the relative importance for all the parameters characterizing the Island model (figure \ref{fig:IS_feat_impo}). We find that all the parameters of the model linked to production, innovation and imitation appear to be relevant for the emergence of sustained economic growth.
	
	The surrogate's performances is presented in Figure \ref{fig:results_IS}, where the first column of the plots refers to the binary outcome setting, while the second one  to the real-valued one. The $F_1$-score displays relatively high values even for low training sample sizes (250 and 500) pointing to a good classification performance of the surrogate (see Figure \ref{fig:IS_f1}). However, it quickly saturates, reaching a plateau around 0.8. Conversely, in the real-valued setting, the surrogate's performance keeps increasing with the training sample size, and it displays remarkably low values of MSE when more than 1000 points are employed (cf. Figure \ref{fig:IS_MSE}).
	
	In both settings, the searching ability of the surrogate behaves in a similar way: the TRP steadily increases with the training sample size (cf. Figures \ref{fig:IS_TPR_class} and \ref{fig:IS_TPR_reg}). In absolute terms, the real-valued setting delivers much better results than the binary one, as for the Brock and Hommes model (section \ref{subsec:BH_results}). In particular, the largest true positive ratio reaches 0.9  for the real-valued case and 0.8 for the binary one. Therefore, by training the surrogate on 2500 points we are able to (i) find 90\% of true positive calibrations (Figure \ref{fig:IS_TPR_reg}) and predict the thickness of the associated distribution of growth rates incurring in a mean squared error of less than 0.08 (Figure \ref{fig:IS_MSE}) using a continuous target variable and, (ii) find 80\% of the true positives (panel \ref{fig:IS_TPR_class}) and correctly classifying around the 80\% of them (panel \ref{fig:IS_f1}) using a binary target variable. 
	
	Given the satisfactory explanatory performance of the surrogate, do we also achieve considerable improvements in the computational time required to perform such exploration exercises? Figures \ref{fig:IS_time_class} and \ref{fig:IS_time_reg} provides a positive answer. Indeed, the surrogate is 3750 times faster than the fully-fledge Island agent-based model. Moreover, the increase in speed is considerably larger than in the Brock and Hommes model. This confirms our intuition on the increasing usefulness of our surrogate modeling approach when the computational cost of the ABM under study is higher. Such a result is a desirable property for real applications, where the complexity of the underlying ABM could even prevent the exploration of the parameter space.

	\subsection{Robustness analysis}\label{sec:IS_stoch}
	
	We now assess the robustness of our training procedure with respect to different surrogate models. More specifically, we compare the XGBoost surrogate employed in the previous analysis with the simpler and more widely used Logit one. Our comparison exercise is performed in a fully stochastic version of the Island agent-based model, where an additional Monte Carlo (MC) is carried out on the seed parameter governing the stochastic terms of the model.
	
 We focus on a binary outcome setting (the one delivering worse performances) and we employ the milder condition that the average growth rate must be positive and sustained, i.e. $AGR>0.5\%$. In this way, the results can be compared to those obtained in the original exercise in \citet{FD03}. We set a budget of 500 evaluations of the ``true'' Islands ABM and run a Monte Carlo exercise of size 100 per parameter combination to generate an MC average of the GDP growth rate that serves as our output variable. Note that this exercise is more complete that the one performed in the previous sections: here, we develop a surrogate model that learns the relationship between parameters and the MC average over their ABM evaluations. Note that this requires many more evaluations of the parameter combination in the true ABM to converge to the statistic required for the label. In our proposed procedure, an MC average growth rate below 0.5\% is labelled ``false'', while AGR above 0.5\% are labeled ``true''. The aim is to learn a surrogate model that accurately classifies parameter combinations as positive or negative calibrations.   
	
 We demonstrate the performance of our active learning approach using two different surrogates: the XGBoost and the faster, less precise, Logit. The former, employed in the analyses carried out in the previous sections, benefits from increased accuracy in exchange for greater computational costs. The latter is a standard statistical model employed regularly for this type of regression analysis. The performance of these alternative surrogates will be evaluated according to the F1-score while training the surrogate, with the final objective of maximizing the precision of the resulting models, i.e. the number of true evaluations which are accurately predicted as positive before they are evaluated. This is a key point to this exercise because real-world use of the proposed approach does not allow us to evaluate all the points in our sample space. Real-world evaluation only provides labels for points that are predicted positive and the resulting performance can only be measured with regard to the true and false positives, with a preference to maximize the former.
	
	The exercise is performed using the Python BOASM package.\footnote{See https://github.com/amirsani/BOASM} The algorithm mirror exactly the one described in Section \ref{sec:surrogate}. The exercise begins by sampling $1000000$ points at random from the Islands parameter space. Given the fixed budget of $500$ evaluations of the true ABM, for both the XGBoost and Logit, the first surrogate is provided with $35$ labelled parameters selected at random from the $1000000$ points, according to the total-variation sampling procedure in \cite{saltelli2010variance}. Then, over several rounds, a surrogate will be fit to the labelled parameters and used to predict a labelling over the $1000000$ points. The predicted labels will then be employed by the proposed procedure to select points that will be added at each round to the set of labelled points. A new surrogate is learned in the subsequent round and the procedure will repeat until the budget of true evaluations has been reached.
	
		\begin{table}[tbp]
			\noindent{\small
				\begin{tabular}{lccccc}
					\toprule
					Surrogate Algorithm & True Negatives & False Positives & False Negatives & True Positives & Precision \\
					\midrule
					Logit & 62 & 22 & 61 & 355 & 94.17\%\\
					XGBoost & 178 & 17 & 0 & 305 & 94.72\%\\
					XGBoost (scaled) & 193 & 2 & 0 & 305 & 99.35\%\\
					\bottomrule
				\end{tabular}
			}
			\caption{Surrogate modelling performance using the learning procedure presented in this paper. Note that only the Precision is computable in a real-life scenario as only True and False Positives are available when positive predicted calibrations are evaluated.}
		\end{table}
	
The proposed procedure results in a comparable precision of 94.17\% and 94.72\% between Logit and XGBoost, respectively. The negligible difference between the precision of the two surrogates suggests that out training procedure provides satisfying results even when the standard Logit statistical model is employed. However, when the XGBoost predicted probabilities are corrected through the Platt scaling procedure,\footnote{Unlike Logit, which produces accurate probabilities for each of the class labels, probabilities produced by non-parametric algorithms such as XGBoost require scaling. Here, we use Platt Scaling to correct the probabilities produced with XGBoost. For more information, see \cite{platt1999probabilistic}.} its precision rises to 99.35\%. Moreover, scaled XGBoost performs is considerably superior to Logit with regard to true vs. false positives. Considering its higher computation costs and need for hyperparameter optimization in using the more precise XGBoost surrogate, users might prefer the faster Logit surrogate when false positives are cheap. Nevertheless, our proposed surrogate modelling procedure works well in both the Logit and XGBoost cases.
	
	\section{Discussion and concluding remarks}\label{sec:conclusions}
	
	In this paper, we have proposed a novel approach to the calibration and parameter space exploration of agent-based models (ABM), which combines the use of supervised machine learning and intelligent sampling to construct a cheap surrogate meta-model. To the best of our knowledge, this is the first attempt to exploit machine-learning techniques for calibration and exploration in an agent-based framework.
	
	Our \textit{machine-learning surrogate} approach is different from kriging, which has been recently applied to ABMs dealing with industrial dynamics \citep{Salle14, Dosi16}, financial networks \citep{Bar16} and macroeconomic issues \citep{Dosietal16,Dosietal17}. In particular, apart from the different statistical framework kriging relies on (it assumes a multivariate Gaussian process), the results it delivers once applied to ABMs may suffer from three relevant limitations. First, kriging is difficult to apply to large scale models, where the number of parameters goes beyond 20. This constrains the modeller to introduce additional procedures to select, a priori, the subset of parameters to study, while leaving the rest constant \citep[see e.g.][]{Dosietal16}. Second, the machine-learning surrogate approach performs better in out-of-sample testing: the typical kriging-based meta-model is tested on 10-20 points within an extremely large space, while our surrogate is tested on samples with size 10000 in the first set of exercises and 1000000 points in the last exercise. Finally, the response surfaces generated by kriging meta-models suffer from smoothness assumptions that collapse interesting patterns, which cannot be captured by common Gaussianity assumptions. This results in incredibly smooth and well-behaved surfaces, which may falsely relate parameters and model behaviour. Given the ragged, unsmooth surfaces commonly reported in agent-based models \citep[see e.g.][]{Gil03, Fab12, Lamperti16}, inferring the behaviour of the true ABM on the basis of the insights produced by kriging meta-model may results in large errors. Further, even when smooth response surfaces exist, kriging requires the selection of the correct prior and kernel. Even in  state-of-the-art likelihood-free approaches, one must select the correct sufficient statistic and acceptance threshold to provide any value.  
	
	The proposed approach manages all these problems.\footnote{In the current work, we also focus on examples dealing with relatively few parameters. This choice is motivated by illustrative reasons and the willingness to use well established models whose code is easily replicable. Further, the results from this paper were produced using a relatively common laptop computer with 16 gigabytes of memory and a 2.4Ghz Intel i7 5500 CPU. The application to a large scale model is currently under development. However, the computational parsimony of the algorithm used to construct our surrogates strongly points to the ability to deal with much richer parameter spaces.} However, the main advantage of our methodology remains in its practical usefulness. Indeed, the surrogate can be learnt at virtually zero computational costs (for research applications) and requires a trivial amount of time to predict areas of the parameter space the modeller should focus on with reasonably good results. Two modelling options are presented, a binary outcome setting and a real-valued one: the first is faster and especially useful when a large number of samples is available, while the second has more explanatory power. Furthermore, the usual trade-off between the quantity of information that needs to be processed (computational costs) and the surrogate performance improvements is, in practice, absent. Ultimately, the surrogate prediction exercises proposed in this paper take less than a minute to complete, with the majority of computation coming from the time to assess the budget of true ABM model evaluations. This means, in practical terms, that the modeller can use an arbitrarily large set of parameter combinations and a relatively small training sample to build the surrogate at almost no cost and leverage the resulting meta-model to gain an insight on the dynamics of the parameter space for further exploration using the original ABM.
	
	Finally, an additional relevant result emerges from the exercises investigated in this paper. The surrogate is much more effective in reducing the relative cost of exploring the properties of the model over the parameter space for the ``Islands'' model, which is more computationally intensive than the Brock and Hommes. This suggests that the adoption of surrogate meta-modelling allows to achieve increasing computational gains as the complexity of the underlying model increases.
	
	This work is only the first step towards a comprehensive assessment of agent-based model properties through machine-learning techniques. Such developments are especially important for complex macroeconomic agent-based models \citep[see e.g.][]{Dosi10,Dosi13,Dosi15,Dosi17,Lilit15} as they could allow the development of a standardized and robust procedure for model calibration and validation, thus closing the existing gap with Dynamics Stochastic General Equilibrium models \citep [see][for a critical comparison of ABM and DSGE models]{Fagiolo_Roventini_2016}. Accordingly, a user-friendly Python surrogate modelling library will also be released for general use.	
	
\subsection*{Acknowledgements}
	\small{A special thank goes to Antoine Mandel, who constantly engaged in fruitful discussions with the authors and provided incredibly valuable insights and suggestions. We would also like to thank Daniele Giachini, Mattia Guerini, Matteo Sostero and Bal\'az K\'egl for their comments.
	Further, we would like to thank all the participants in seminars and workshops held at Scuola Superiore Sant'Anna (Pisa), the XXI WEHIA conference (Castellon), the XXII CEF conference (Bordeaux), NIPS 2016 "What If? Inference and Learning of Hypothetical and Counterfactual Interventions in Complex Systems" workshop (Barcelona), CCS 2016 conference (Amsterdam) and 2016 Paris-Bielefeld Workshop on Agent-Based Modeling (Paris). FL acknowledges financial support from European Union's FP7 project IMPRESSIONS (G.A. No 603416). AS acknowledges financial support from the H2020 project DOLFINS (G.A. No 640772) and hardware support from NVIDIA Corporation, AdapData SAS and the Grid5000 testbed for this research. AR acknowledges financial support from European Union’s FP7 IMPRESSIONS, H2020 DOLFINS  and H2020 ISIGROWTH (G.A. No 649186) projects.}
	
	\bigskip
	
	\singlespacing
	\footnotesize

	\bibliographystyle{apa}
	\bibliography{bib_abmopt}

\begin{thebibliography}{}

\bibitem[\protect\astroncite{Alfarano et~al.}{2005}]{Alf05}
Alfarano, S., Lux, T., and Wagner, F. (2005).
\newblock Estimation of agent-based models: The case of an asymmetric herding
  model.
\newblock {\em Computational Economics}, 26(1):19--49.

\bibitem[\protect\astroncite{Alfarano et~al.}{2006}]{Alf06}
Alfarano, S., Lux, T., and Wagner, F. (2006).
\newblock Estimation of a simple agent-based model of financial markets: An
  application to australian stock and foreign exchange data.
\newblock {\em Physica A: Statistical Mechanics and its Applications},
  370(1):38 -- 42.

\bibitem[\protect\astroncite{Amilon}{2008}]{Ami08}
Amilon, H. (2008).
\newblock Estimation of an adaptive stock market model with heterogeneous
  agents.
\newblock {\em Journal of Empirical Finance}, 15(2):342 -- 362.

\bibitem[\protect\astroncite{An and Wilensky}{2009}]{An2009}
An, G. and Wilensky, U. (2009).
\newblock From artificial life to in silico medicine.
\newblock In Komosinski, M. and Adamatzky, A., editors, {\em Artificial Life
  Models in Software}, pages 183--214. Springer London, London.

\bibitem[\protect\astroncite{Anderson et~al.}{1972}]{And72}
Anderson, P.~W. et~al. (1972).
\newblock More is different.
\newblock {\em Science}, 177(4047):393--396.

\bibitem[\protect\astroncite{Archer and Kimes}{2008}]{archer2008empirical}
Archer, K.~J. and Kimes, R.~V. (2008).
\newblock Empirical characterization of random forest variable importance
  measures.
\newblock {\em Computational Statistics \& Data Analysis}, 52(4):2249--2260.

\bibitem[\protect\astroncite{Assenza et~al.}{2015}]{Assenza2015}
Assenza, T., Gatti, D.~D., and Grazzini, J. (2015).
\newblock Emergent dynamics of a macroeconomic agent based model with capital
  and credit.
\newblock {\em Journal of Economic Dynamics and Control}, 50:5--28.

\bibitem[\protect\astroncite{Banerjee}{1992}]{Ban92}
Banerjee, A.~V. (1992).
\newblock A simple model of herd behavior.
\newblock {\em The Quarterly Journal of Economics}, 107(3):797--817.

\bibitem[\protect\astroncite{Barde}{2016a}]{Barde16}
Barde, S. (2016a).
\newblock Direct comparison of agent-based models of herding in financial
  markets.
\newblock {\em Journal of Economic Dynamics and Control}, 73:329 -- 353.

\bibitem[\protect\astroncite{Barde}{2016b}]{Bar15}
Barde, S. (2016b).
\newblock A practical, accurate, information criterion for nth order markov
  processes.
\newblock {\em Computational Economics}, pages 1--44.

\bibitem[\protect\astroncite{Bargigli et~al.}{2016}]{Bar16}
Bargigli, L., Riccetti, L., Russo, A., and Gallegati, M. (2016).
\newblock {Network Calibration and Metamodeling of a Financial Accelerator
  Agent Based Model}.
\newblock Working papers, economics, Universit\'a degli Studi di Firenze,
  Dipartimento di Scienze per l'Economia e l'Impresa.

\bibitem[\protect\astroncite{Bergstra and Bengio}{2012}]{bergstra2012random}
Bergstra, J. and Bengio, Y. (2012).
\newblock Random search for hyper-parameter optimization.
\newblock {\em Journal of Machine Learning Research}, 13(Feb):281--305.

\bibitem[\protect\astroncite{Booker et~al.}{1999}]{Book99}
Booker, A., Dennis, J.E., J., Frank, P., Serafini, D., Torczon, V., and
  Trosset, M. (1999).
\newblock A rigorous framework for optimization of expensive functions by
  surrogates.
\newblock {\em Structural optimization}, 17(1):1--13.

\bibitem[\protect\astroncite{Boswijk et~al.}{2007}]{Bos07}
Boswijk, H., Hommes, C., and Manzan, S. (2007).
\newblock Behavioral heterogeneity in stock prices.
\newblock {\em Journal of Economic Dynamics and Control}, 31(6):1938 -- 1970.

\bibitem[\protect\astroncite{Bottazzi and Secchi}{2006}]{Bose}
Bottazzi, G. and Secchi, A. (2006).
\newblock Explaining the distribution of firm growth rates.
\newblock {\em The RAND Journal of Economics}, 37(2):235--256.

\bibitem[\protect\astroncite{Breiman}{2001}]{breiman2001random}
Breiman, L. (2001).
\newblock Random forests.
\newblock {\em Machine learning}, 45(1):5--32.

\bibitem[\protect\astroncite{Breiman et~al.}{1984}]{breiman1984classification}
Breiman, L., Friedman, J., Stone, C.~J., and Olshen, R.~A. (1984).
\newblock {\em Classification and regression trees}.
\newblock CRC press.

\bibitem[\protect\astroncite{Brock and Hommes}{1997}]{BH97}
Brock, W.~A. and Hommes, C.~H. (1997).
\newblock A rational route to randomness.
\newblock {\em Econometrica}, 65(5):1059--1095.

\bibitem[\protect\astroncite{Brock and Hommes}{1998}]{BH98}
Brock, W.~A. and Hommes, C.~H. (1998).
\newblock Heterogeneous beliefs and routes to chaos in a simple asset pricing
  model.
\newblock {\em Journal of Economic Dynamics and Control}, 22(8--9):1235 --
  1274.

\bibitem[\protect\astroncite{Brown et~al.}{2005}]{brown2005}
Brown, D.~G., Page, S., Riolo, R., Zellner, M., and Rand, W. (2005).
\newblock Path dependence and the validation of agent-based spatial models of
  land use.
\newblock {\em International Journal of Geographical Information Science},
  19(2):153--174.

\bibitem[\protect\astroncite{Caiani et~al.}{2016}]{Caiani2016}
Caiani, A., Godin, A., Caverzasi, E., Gallegati, M., Kinsella, S., and
  Stiglitz, J.~E. (2016).
\newblock Agent based-stock flow consistent macroeconomics: Towards a benchmark
  model.
\newblock {\em Journal of Economic Dynamics and Control}, 69:375--408.

\bibitem[\protect\astroncite{Carley et~al.}{2006}]{carley2006}
Carley, K.~M., Fridsma, D.~B., Casman, E., Yahja, A., Altman, N., Chen, L.-C.,
  Kaminsky, B., and Nave, D. (2006).
\newblock Biowar: scalable agent-based model of bioattacks.
\newblock {\em IEEE Transactions on Systems, Man, and Cybernetics-Part A:
  Systems and Humans}, 36(2):252--265.

\bibitem[\protect\astroncite{Castaldi and Dosi}{2009}]{CD09}
Castaldi, C. and Dosi, G. (2009).
\newblock The patterns of output growth of firms and countries: Scale
  invariances and scale specificities.
\newblock {\em Empirical Economics}, 37(3):475--495.

\bibitem[\protect\astroncite{Chen et~al.}{2012}]{Che12}
Chen, S.-H., Chang, C.-L., and Du, Y.-R. (2012).
\newblock Agent-based economic models and econometrics.
\newblock {\em The Knowledge Engineering Review}, 27:187--219.

\bibitem[\protect\astroncite{Chen and Guestrin}{2016}]{chen2016xgboost}
Chen, T. and Guestrin, C. (2016).
\newblock Xgboost: A scalable tree boosting system.
\newblock In {\em Proceedings of the 22Nd ACM SIGKDD International Conference
  on Knowledge Discovery and Data Mining}, pages 785--794. ACM.

\bibitem[\protect\astroncite{Chiarella et~al.}{2009}]{Chi09}
Chiarella, C., Iori, G., and Perell{\'o}, J. (2009).
\newblock The impact of heterogeneous trading rules on the limit order book and
  order flows.
\newblock {\em Journal of Economic Dynamics and Control}, 33(3):525--537.

\bibitem[\protect\astroncite{Cire{\c{s}}an et~al.}{2013}]{cire}
Cire{\c{s}}an, D.~C., Giusti, A., Gambardella, L.~M., and Schmidhuber, J.
  (2013).
\newblock Mitosis detection in breast cancer histology images with deep neural
  networks.
\newblock In {\em International Conference on Medical Image Computing and
  Computer-assisted Intervention}, pages 411--418. Springer.

\bibitem[\protect\astroncite{Claesen et~al.}{2014}]{claesen2014easy}
Claesen, M., Simm, J., Popovic, D., Moreau, Y., and De~Moor, B. (2014).
\newblock Easy hyperparameter search using optunity.
\newblock {\em arXiv preprint arXiv:1412.1114}.

\bibitem[\protect\astroncite{Conti and O'Hagan}{2010}]{conti2010bayesian}
Conti, S. and O'Hagan, A. (2010).
\newblock Bayesian emulation of complex multi-output and dynamic computer
  models.
\newblock {\em Journal of statistical planning and inference}, 140(3):640--651.

\bibitem[\protect\astroncite{Dawid et~al.}{2014a}]{Dawid14}
Dawid, H., Gemkow, S., Harting, P., Van~der Hoog, S., and Neugart, M. (2014a).
\newblock Agent-based macroeconomic modeling and policy analysis: the eurace@
  unibi model.
\newblock Technical report, Bielefeld Working Papers in Economics and
  Management.

\bibitem[\protect\astroncite{Dawid et~al.}{2014b}]{dawid2014economic}
Dawid, H., Harting, P., and Neugart, M. (2014b).
\newblock Economic convergence: Policy implications from a heterogeneous agent
  model.
\newblock {\em Journal of Economic Dynamics and Control}, 44:54--80.

\bibitem[\protect\astroncite{De~Marchi}{2005}]{Dem05}
De~Marchi, S. (2005).
\newblock {\em Computational and mathematical modeling in the social sciences}.
\newblock Cambridge University Press.

\bibitem[\protect\astroncite{Dosi}{1988}]{Dos88}
Dosi, G. (1988).
\newblock Sources, procedures and microeconomic effects of innovation.
\newblock {\em Journal of Economic Literature}, 26:126--71.

\bibitem[\protect\astroncite{Dosi et~al.}{2013}]{Dosi13}
Dosi, G., Fagiolo, G., Napoletano, M., and Roventini, A. (2013).
\newblock {Income distribution, credit and fiscal policies in an agent-based
  Keynesian model}.
\newblock {\em Journal of Economic Dynamics and Control}, 37(8):1598--1625.

\bibitem[\protect\astroncite{Dosi et~al.}{2015}]{Dosi15}
Dosi, G., Fagiolo, G., Napoletano, M., Roventini, A., and Treibich, T. (2015).
\newblock {Fiscal and monetary policies in complex evolving economies}.
\newblock {\em Journal of Economic Dynamics and Control}, 52(C):166--189.

\bibitem[\protect\astroncite{Dosi et~al.}{2010}]{Dosi10}
Dosi, G., Fagiolo, G., and Roventini, A. (2010).
\newblock {Schumpeter meeting Keynes: A policy-friendly model of endogenous
  growth and business cycles}.
\newblock {\em Journal of Economic Dynamics and Control}, 34(9):1748--1767.

\bibitem[\protect\astroncite{Dosi et~al.}{2017a}]{Dosi17}
Dosi, G., Pereira, M., Roventini, A., and Virgillito, M. (2017a).
\newblock When more flexibility yields more fragility: The microfoundations of
  keynesian aggregate unemployment.
\newblock {\em Journal of Economic Dynamics and Control}, forthcoming.

\bibitem[\protect\astroncite{Dosi et~al.}{2016}]{Dosietal16}
Dosi, G., Pereira, M., Roventini, A., and Virgillito, M.~E. (2016).
\newblock {The Effects of Labour Market Reforms upon Unemployment and Income
  Inequalities: an Agent Based Model}.
\newblock LEM Working Papers Series 2016-27, Scuola Superiore Sant'Anna.

\bibitem[\protect\astroncite{Dosi et~al.}{2017b}]{Dosietal17}
Dosi, G., Pereira, M., Roventini, A., and Virgillito, M.~E. (2017b).
\newblock Causes and consequences of hysteresis: Aggregate demand, productivity
  and employment.
\newblock LEM Working Papers Series 2017-07, Scuola Superiore Sant'Anna.

\bibitem[\protect\astroncite{Dosi et~al.}{2017c}]{Dosi16}
Dosi, G., Pereira, M.~C., and Virgillito, M.~E. (2017c).
\newblock On the robustness of the fat-tailed distribution of firm growth
  rates: a global sensitivity analysis.
\newblock {\em Journal of Economic Interaction and Coordination}, pages 1--21.

\bibitem[\protect\astroncite{Effken et~al.}{2012}]{eff12}
Effken, J.~A., Carley, K.~M., Lee, J.-S., Brewer, B.~B., and Verran, J.~A.
  (2012).
\newblock Simulating nursing unit performance with orgahead: strengths and
  challenges.
\newblock {\em Computers, informatics, nursing: CIN}, 30(11):620.

\bibitem[\protect\astroncite{Fabretti}{2012}]{Fab12}
Fabretti, A. (2012).
\newblock On the problem of calibrating an agent based model for financial
  markets.
\newblock {\em Journal of Economic Interaction and Coordination},
  8(2):277--293.

\bibitem[\protect\astroncite{Fagiolo et~al.}{2007}]{Fagiolo07}
Fagiolo, G., Birchenhall, C., and Windrum, P. (2007).
\newblock Empirical validation in agent-based models: Introduction to the
  special issue.
\newblock {\em Computational Economics}, 30(3):189--194.

\bibitem[\protect\astroncite{Fagiolo and Dosi}{2003}]{FD03}
Fagiolo, G. and Dosi, G. (2003).
\newblock {Exploitation, exploration and innovation in a model of endogenous
  growth with locally interacting agents}.
\newblock {\em Structural Change and Economic Dynamics}, 14(3):237--273.

\bibitem[\protect\astroncite{Fagiolo et~al.}{2008}]{FNR08}
Fagiolo, G., Napoletano, M., and Roventini, A. (2008).
\newblock Are output growth-rate distributions fat-tailed? some evidence from
  oecd countries.
\newblock {\em Journal of Applied Econometrics}, 23(5):639--669.

\bibitem[\protect\astroncite{Fagiolo and
  Roventini}{2012}]{Fagiolo_Roventini_2012}
Fagiolo, G. and Roventini, A. (2012).
\newblock Macroeconomic policy in dsge and agent-based models.
\newblock {\em Revue de l'OFCE}, 124:67--116.

\bibitem[\protect\astroncite{Fagiolo and
  Roventini}{2017}]{Fagiolo_Roventini_2016}
Fagiolo, G. and Roventini, A. (2017).
\newblock Macroeconomic policy in dsge and agent-based models redux: New
  developments and challenges ahead.
\newblock {\em Journal of Artificial Societies and Social Simulation}, 20(1).

\bibitem[\protect\astroncite{Feurer et~al.}{2015}]{feurer2015efficient}
Feurer, M., Klein, A., Eggensperger, K., Springenberg, J., Blum, M., and
  Hutter, F. (2015).
\newblock Efficient and robust automated machine learning.
\newblock In {\em Advances in Neural Information Processing Systems}, pages
  2962--2970.

\bibitem[\protect\astroncite{Franke}{2009}]{Fra09}
Franke, R. (2009).
\newblock Applying the method of simulated moments to estimate a small
  agent-based asset pricing model.
\newblock {\em Journal of Empirical Finance}, 16(5):804 -- 815.

\bibitem[\protect\astroncite{Franke and Westerhoff}{2012}]{franke2012}
Franke, R. and Westerhoff, F. (2012).
\newblock Structural stochastic volatility in asset pricing dynamics:
  Estimation and model contest.
\newblock {\em Journal of Economic Dynamics and Control}, 36(8):1193--1211.

\bibitem[\protect\astroncite{Freund}{1990}]{freund1990boosting}
Freund, Y. (1990).
\newblock Boosting a weak learning algorithm by majority.
\newblock In {\em COLT}, volume~90, pages 202--216.

\bibitem[\protect\astroncite{Freund et~al.}{1996}]{freund1996experiments}
Freund, Y., Schapire, R.~E., et~al. (1996).
\newblock Experiments with a new boosting algorithm.
\newblock In {\em Icml}, volume~96, pages 148--156.

\bibitem[\protect\astroncite{Gallegati and Kirman}{2012}]{GK12}
Gallegati, M. and Kirman, A. (2012).
\newblock Reconstructing economics.
\newblock {\em Complexity Economics}, 1(1):5--31.

\bibitem[\protect\astroncite{Gilli and Winker}{2003}]{Gil03}
Gilli, M. and Winker, P. (2003).
\newblock A global optimization heuristic for estimating agent based models.
\newblock {\em Computational Statistics \& Data Analysis}, 42(3):299 -- 312.
\newblock Computational Ecomometrics.

\bibitem[\protect\astroncite{Goldberg et~al.}{2011}]{goldberg2011oasis}
Goldberg, A.~B., Zhu, X., Furger, A., and Xu, J.-M. (2011).
\newblock Oasis: Online active semi-supervised learning.
\newblock In {\em AAAI}.

\bibitem[\protect\astroncite{Grazzini}{2012}]{Graz12}
Grazzini, J. (2012).
\newblock Analysis of the emergent properties: Stationarity and ergodicity.
\newblock {\em Journal of Artificial Societies and Social Simulation}, 15(2):7.

\bibitem[\protect\astroncite{Grazzini and Richiardi}{2015}]{Graz15}
Grazzini, J. and Richiardi, M. (2015).
\newblock Estimation of ergodic agent-based models by simulated minimum
  distance.
\newblock {\em Journal of Economic Dynamics and Control}, 51:148 -- 165.

\bibitem[\protect\astroncite{Grazzini et~al.}{2017}]{Rich15}
Grazzini, J., Richiardi, M.~G., and Tsionas, M. (2017).
\newblock Bayesian estimation of agent-based models.
\newblock {\em Journal of Economic Dynamics and Control}, 77:26 -- 47.

\bibitem[\protect\astroncite{Grimm and Railsback}{2013}]{grimm2013}
Grimm, V. and Railsback, S.~F. (2013).
\newblock {\em Individual-based modeling and ecology}.
\newblock Princeton university press.

\bibitem[\protect\astroncite{Guerini and Moneta}{2016}]{Gue16}
Guerini, M. and Moneta, A. (2016).
\newblock {A Method for Agent-Based Models Validation}.
\newblock LEM Papers Series 2016/16, Laboratory of Economics and Management
  (LEM), Sant'Anna School of Advanced Studies, Pisa, Italy.

\bibitem[\protect\astroncite{Herlands et~al.}{2015}]{herlands2015scalable}
Herlands, W., Wilson, A., Nickisch, H., Flaxman, S., Neill, D., Van~Panhuis,
  W., and Xing, E. (2015).
\newblock Scalable gaussian processes for characterizing multidimensional
  change surfaces.
\newblock {\em arXiv preprint arXiv:1511.04408}.

\bibitem[\protect\astroncite{Ilachinski}{1997}]{ila97}
Ilachinski, A. (1997).
\newblock Irreducible semi-autonomous adaptive combat (isaac): An
  artificial-life approach to land warfare.
\newblock Technical report, DTIC Document.

\bibitem[\protect\astroncite{Kukacka and Barunik}{2016}]{Kuk16}
Kukacka, J. and Barunik, J. (2016).
\newblock Estimation of financial agent-based models with simulated maximum
  likelihood.
\newblock IES Working Paper 7/2016, Charles University of Prague.

\bibitem[\protect\astroncite{Lamperti}{2016}]{Lamperti16}
Lamperti, F. (2016).
\newblock {Empirical Validation of Simulated Models through the GSL-div: an
  Illustrative Application}.
\newblock LEM Papers Series 2016/18, Laboratory of Economics and Management
  (LEM), Sant'Anna School of Advanced Studies, Pisa, Italy.

\bibitem[\protect\astroncite{Lamperti}{2017}]{Lamperti15}
Lamperti, F. (2017).
\newblock An information theoretic criterion for empirical validation of
  simulation models.
\newblock {\em Econometrics and Statistics}, forthcoming.

\bibitem[\protect\astroncite{Lamperti et~al.}{2017}]{DSK}
Lamperti, F., Dosi, G., Napoletano, M., Roventini, A., and Sapio, A. (2017).
\newblock Faraway, so close: coupled climate and economic dynamics in an agent
  based integrated assessment model.
\newblock Lem working papers series, Scuola Superiore Sant'Anna.

\bibitem[\protect\astroncite{Lamperti and Mattei}{2016}]{LM16}
Lamperti, F. and Mattei, C.~E. (2016).
\newblock {Going Up and Down: Rethinking the Empirics of Growth in the
  Developing and Newly Industrialized World}.
\newblock LEM Papers Series 2016/01, Laboratory of Economics and Management
  (LEM), Sant'Anna School of Advanced Studies, Pisa, Italy.

\bibitem[\protect\astroncite{Leal et~al.}{2014}]{HFT15}
Leal, S.~J., Napoletano, M., Roventini, A., and Fagiolo, G. (2014).
\newblock Rock around the clock: an agent-based model of low-and high-frequency
  trading.
\newblock {\em Journal of Evolutionary Economics}, pages 1--28.

\bibitem[\protect\astroncite{Lee et~al.}{2015}]{lee15}
Lee, J.-S., Filatova, T., Ligmann-Zielinska, A., Hassani-Mahmooei, B.,
  Stonedahl, F., Lorscheid, I., Voinov, A., Polhill, J.~G., Sun, Z., and
  Parker, D.~C. (2015).
\newblock The complexities of agent-based modeling output analysis.
\newblock {\em Journal of Artificial Societies and Social Simulation}, 18(4):4.

\bibitem[\protect\astroncite{Li et~al.}{2013}]{li2013benchmark}
Li, X., Engelbrecht, A., and Epitropakis, M.~G. (2013).
\newblock Benchmark functions for cec'2013 special session and competition on
  niching methods for multimodal function optimization.
\newblock {\em RMIT University, Evolutionary Computation and Machine Learning
  Group, Australia, Tech. Rep}.

\bibitem[\protect\astroncite{Louppe et~al.}{2013}]{louppe2013understanding}
Louppe, G., Wehenkel, L., Sutera, A., and Geurts, P. (2013).
\newblock Understanding variable importances in forests of randomized trees.
\newblock In {\em Advances in neural information processing systems}, pages
  431--439.

\bibitem[\protect\astroncite{Lux and Marchesi}{2000}]{Lux00}
Lux, T. and Marchesi, M. (2000).
\newblock Volatility clustering in financial markets: a microsimulation of
  interacting agents.
\newblock {\em International journal of theoretical and applied finance},
  3(04):675--702.

\bibitem[\protect\astroncite{Macy and Willer}{2002}]{macy2002}
Macy, M.~W. and Willer, R. (2002).
\newblock From factors to actors: Computational sociology and agent-based
  modeling.
\newblock {\em Annual review of sociology}, pages 143--166.

\bibitem[\protect\astroncite{Marks}{2013}]{Marks13}
Marks, R.~E. (2013).
\newblock Validation and model selection: Three similarity measures compared.
\newblock {\em Complexity Economics}, 2(1):41--61.

\bibitem[\protect\astroncite{Morokoff and Caflisch}{1994}]{morokoff1994quasi}
Morokoff, W.~J. and Caflisch, R.~E. (1994).
\newblock Quasi-random sequences and their discrepancies.
\newblock {\em SIAM Journal on Scientific Computing}, 15(6):1251--1279.

\bibitem[\protect\astroncite{Moss}{2008}]{moss2008alternative}
Moss, S. (2008).
\newblock Alternative approaches to the empirical validation of agent-based
  models.
\newblock {\em Journal of Artificial Societies and Social Simulation}, 11(1):5.

\bibitem[\protect\astroncite{Petrovic et~al.}{2011}]{petro}
Petrovic, S., Osborne, M., and Lavrenko, V. (2011).
\newblock Rt to win! predicting message propagation in twitter.
\newblock {\em ICWSM}, 11:586--589.

\bibitem[\protect\astroncite{Platt et~al.}{1999}]{platt1999probabilistic}
Platt, J. et~al. (1999).
\newblock Probabilistic outputs for support vector machines and comparisons to
  regularized likelihood methods.
\newblock {\em Advances in large margin classifiers}, 10(3):61--74.

\bibitem[\protect\astroncite{Popoyan et~al.}{2017}]{Lilit15}
Popoyan, L., Napoletano, M., and Roventini, A. (2017).
\newblock {Taming Macroeconomic Instability: Monetary and Macro Prudential
  Policy Interactions in an Agent-Based Model}.
\newblock {\em Journal of Economic Behavior \& Organization}, 134:117--140.

\bibitem[\protect\astroncite{Rasmussen and
  Williams}{2006}]{rasmussen2006gaussian}
Rasmussen, C.~E. and Williams, C. K.~I. (2006).
\newblock {\em Gaussian processes for machine learning}.
\newblock MIT Press.

\bibitem[\protect\astroncite{Recchioni et~al.}{2015}]{Rec15}
Recchioni, M.~C., Tedeschi, G., and Gallegati, M. (2015).
\newblock A calibration procedure for analyzing stock price dynamics in an
  agent-based framework.
\newblock {\em Journal of Economic Dynamics and Control}, 60:1 -- 25.

\bibitem[\protect\astroncite{Ross et~al.}{2011}]{ross2011reduction}
Ross, S., Gordon, G.~J., and Bagnell, D. (2011).
\newblock A reduction of imitation learning and structured prediction to
  no-regret online learning.
\newblock In {\em AISTATS}, volume 1(2), page~6.

\bibitem[\protect\astroncite{Ryabko}{2016}]{ryabko2016things}
Ryabko, D. (2016).
\newblock Things bayes can't do.
\newblock In {\em International Conference on Algorithmic Learning Theory},
  pages 253--260. Springer.

\bibitem[\protect\astroncite{Salle and Yildizoglu}{2014}]{Salle14}
Salle, I. and Yildizoglu, M. (2014).
\newblock {Efficient Sampling and Meta-Modeling for Computational Economic
  Models}.
\newblock {\em Computational Economics}, 44(4):507--536.

\bibitem[\protect\astroncite{Saltelli et~al.}{2010}]{saltelli2010variance}
Saltelli, A., Annoni, P., Azzini, I., Campolongo, F., Ratto, M., and Tarantola,
  S. (2010).
\newblock Variance based sensitivity analysis of model output. design and
  estimator for the total sensitivity index.
\newblock {\em Computer Physics Communications}, 181(2):259--270.

\bibitem[\protect\astroncite{Settles}{2010}]{settles2010active}
Settles, B. (2010).
\newblock Active learning literature survey.
\newblock Technical Report 55-66, University of Wisconsin, Madison.

\bibitem[\protect\astroncite{Squazzoni}{2010}]{squaz10}
Squazzoni, F. (2010).
\newblock The impact of agent-based models in the social sciences after 15
  years of incursions.
\newblock {\em History of Economic Ideas}, pages 197--233.

\bibitem[\protect\astroncite{Subbotin}{1923}]{subbo}
Subbotin, M.~T. (1923).
\newblock On the law of frequency of error.
\newblock {\em Matematicheskii Sbornik}, 31(2):296--301.

\bibitem[\protect\astroncite{ten Broeke et~al.}{2016}]{ten2016sensitivity}
ten Broeke, G., van Voorn, G., and Ligtenberg, A. (2016).
\newblock Which sensitivity analysis method should i use for my agent-based
  model?
\newblock {\em Journal of Artificial Societies \& Social Simulation}, 19(1).

\bibitem[\protect\astroncite{Tesfatsion and Judd}{2006}]{TJ06}
Tesfatsion, L. and Judd, K.~L. (2006).
\newblock {\em Handbook of computational economics: agent-based computational
  economics}, volume~2.
\newblock Elsevier.

\bibitem[\protect\astroncite{Thiele et~al.}{2014}]{thiele2014facilitating}
Thiele, J.~C., Kurth, W., and Grimm, V. (2014).
\newblock Facilitating parameter estimation and sensitivity analysis of
  agent-based models: A cookbook using netlogo and r.
\newblock {\em Journal of Artificial Societies and Social Simulation},
  17(3):11.

\bibitem[\protect\astroncite{van~der Hoog}{2016}]{Sander}
van~der Hoog, S. (2016).
\newblock {Deep Learning in Agent-Based Models: A Prospectus}.
\newblock Technical report, Faculty of Business Administration and Economics,
  Bielefeld University.

\bibitem[\protect\astroncite{Van~Rijsbergen}{1979}]{Van79}
Van~Rijsbergen, C. (1979).
\newblock {\em Information Retrieval}.
\newblock London: Butterworths.

\bibitem[\protect\astroncite{Weeks}{1995}]{Week95}
Weeks, M. (1995).
\newblock Circumventing the curse of dimensionality in applied work using
  computer intensive methods.
\newblock {\em The Economic Journal}, 105(429):520--530.

\bibitem[\protect\astroncite{Wilson et~al.}{2015}]{wilson2015thoughts}
Wilson, A.~G., Dann, C., and Nickisch, H. (2015).
\newblock Thoughts on massively scalable gaussian processes.
\newblock {\em arXiv preprint arXiv:1511.01870}.

\bibitem[\protect\astroncite{Winker et~al.}{2007}]{Win07}
Winker, P., Gilli, M., and Jeleskovic, V. (2007).
\newblock An objective function for simulation based inference on exchange rate
  data.
\newblock {\em Journal of Economic Interaction and Coordination},
  2(2):125--145.

\bibitem[\protect\astroncite{Wolpert}{2002}]{wolpert2002supervised}
Wolpert, D.~H. (2002).
\newblock The supervised learning no-free-lunch theorems.
\newblock In {\em Soft Computing and Industry}, pages 25--42. Springer.

\bibitem[\protect\astroncite{Wong}{2015}]{wong2015evolutionary}
Wong, K.-C. (2015).
\newblock Evolutionary multimodal optimization: A short survey.
\newblock {\em arXiv preprint arXiv:1508.00457}.

\bibitem[\protect\astroncite{Zhu}{2005}]{zhu2005semi}
Zhu, X. (2005).
\newblock Semi-supervised learning literature survey.
\newblock Technical report, University of Wisconsin-Madison.

\end{thebibliography}
	
\end{document}